\documentclass{jfm}
\usepackage{graphicx}
\usepackage{multirow}
\usepackage{amsmath}
\usepackage{amssymb}
\usepackage{float}
\usepackage{epstopdf, epsfig}
\usepackage{xcolor}
\usepackage{tikz}
\usepackage{mathabx}
\usepackage{caption}
\definecolor{myred}{RGB}{192, 0, 0}
\definecolor{mycyan}{RGB}{0, 112, 192}
\definecolor{myblue}{RGB}{0, 0, 255}

\shorttitle{Flow pulsations in high-speed double cones}
\shortauthor{S. Das and S. Duvvuri}

\title{A model for pulsation in\\high-speed double cone flow}

\author{Subhajit Das
 \and Subrahmanyam Duvvuri\corresp{\email{subrahmanyam@iisc.ac.in}}}

\affiliation{Turbulent Shear Flow Physics and Engineering Laboratory\\Department of Aerospace Engineering, Indian Institute of Science, Bengaluru 560 012}

\begin{document}

\maketitle

\begin{abstract}
Periodic large-scale shock-wave unsteadiness over a canonical double cone, termed in literature as ``pulsation,'' is experimentally studied at Mach 6. The general double cone geometry is defined by three non-dimensional geometric parameters: fore- and aft-cone angles ($\theta_1$ and $\theta_2$), and ratio of the conical slant lengths ($\mathit{\Lambda}$). While existing literature on pulsation offers detailed qualitative and phenomenological discussions, it is seen that analytical approaches to obtain insight into the unsteady flow phenomena are missing. The present effort is aimed at addressing this gap. Self-sustained flow pulsations for a particular double cone configuration with $\theta_1 = 0^\circ$ and $\theta_2 = 90^\circ$, commonly referred to as the spike-cylinder, is investigated in the $\mathit{\Lambda}$ parameter space. High-speed schlieren imaging and time-resolved pressure measurements are performed in the unsteady flow. The non-dimensional pulsation frequency (Strouhal number) is observed to increase monotonically with $\mathit{\Lambda}$. Schlieren and pressure data suggest that the unsteadiness is driven by a cyclic process involving the formation of high-pressure gas near the aft-cone and its subsequent expansion through the separation region formed over the fore-cone. Building on this understanding, a detailed analytical model for the flow is developed with no empirical parameters. The model successfully predicts the experimentally-measured Strouhal number, and provides an in-depth understanding of the mechanisms that drive flow pulsations.
\end{abstract}


\section{Introduction}
\label{Introduction}
Separated flows in the high-speed regime (supersonic/hypersonic) are of active research interest due to their scientific importance and also their direct relevance to high-speed flight. Temporal unsteadiness is often inherent to these flows, primarily driven by the interactions between shock wave and shear layer/separation regions. Within the broader class of flows involving shock wave and shear layer interactions, a subset known as shock-wave/boundary-layer interaction (SBLI) has been widely studied \citep[see, for instance,][and references therein]{beresh2002relationship,touber2011low,Gaitonde2015,murugan2016shock}. In a typical SBLI scenario, an external shock wave impinges on a boundary layer flow and thereby induces a strong adverse pressure gradient in the wall-parallel direction. This leads to flow separation from the wall, with the separation region typically remaining localized. However, in certain flow situations, due to constraints imposed by the flow geometry (\textit{e.g.}, a large flow turning angle around a corner), the adverse pressure gradient can be sufficiently large to cause the separation region to grow in an unsteady manner. In such cases the separation regions can grow to dimensions comparable to the geometric length scales governing the flow. This phenomenon defines a distinct class of problems where the nature of interaction between shock waves and large separation regions is fundamentally different from conventional SBLI, and is categorized as shock wave/separation region interaction, referred to in short as SSRI \citep{duvvuri2023shock}. \par

\begin{figure}
  \centering
  \includegraphics[width=0.6\columnwidth]{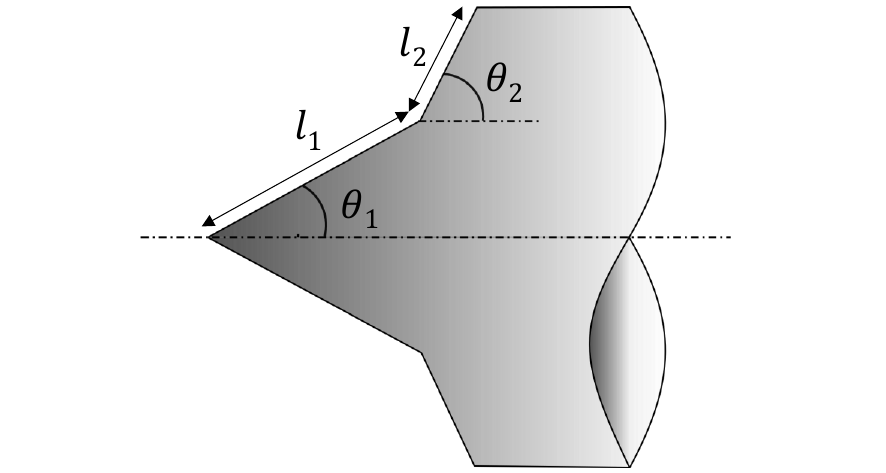} 
  \captionsetup{width=1\textwidth} 
  \caption{A schematic of the double-cone geometry. Fore-cone half-angle $\theta_1$, aft-cone half-angle $\theta_2$ and slant length ratio $\mathit{\Lambda} = l_2/l_1$ are three non-dimensional parameters that completely define the geometry.}
  \label{fig:1}
\end{figure}

The high-speed double cone presents a canonical example of SSRI. It offers a well-defined geometric configuration for detailed study of the flow dynamics associated with SSRI. The double cone model comprises two adjacent right circular conical sections along a
single axis (see figure~\ref{fig:1}). The model geometry is described by the two cone half-angles $\theta_1$, $\theta_2$ and slant lengths $l_1$, $l_2$ of the conical sections. The double cone configuration can be fully defined by three non-dimensional parameters, taken here to be
$\theta_1$, $\theta_2$ and $\mathit{\Lambda}$ = $l_2/l_1$. A special case of the general double cone configuration is the spiked cylinder, where $\theta_1$ and $\theta_2$ are fixed at $\theta_1$ = $0^\circ$ and $\theta_1$ = $90^\circ$, respectively. The spike-cylinder geometry is defined by a single non-dimensional parameter, \textit{i.e.}, $\mathit{\Lambda}$. High-speed flow over the spike-cylinder has attracted significant research attention since the 1950s (see \citeauthor{sahoo2021shock} \citeyear{sahoo2021shock}, \citeauthor{duvvuri2023shock} \citeyear{duvvuri2023shock} and references therein). Early experimental investigations clearly revealed the presence of unsteady and periodic flow phenomena \citep{Mair01071952,maull1960hypersonic,wood1962hypersonic,Kenworthy1975,Kenworthy1978}. For a specific certain range of $\mathit{\Lambda}$, the spatial extent of these unsteady motions can become much larger than those typically observed in conventional SBLI scenarios. \par
\begin{figure}
  \centerline{\includegraphics[width=1.05\columnwidth]{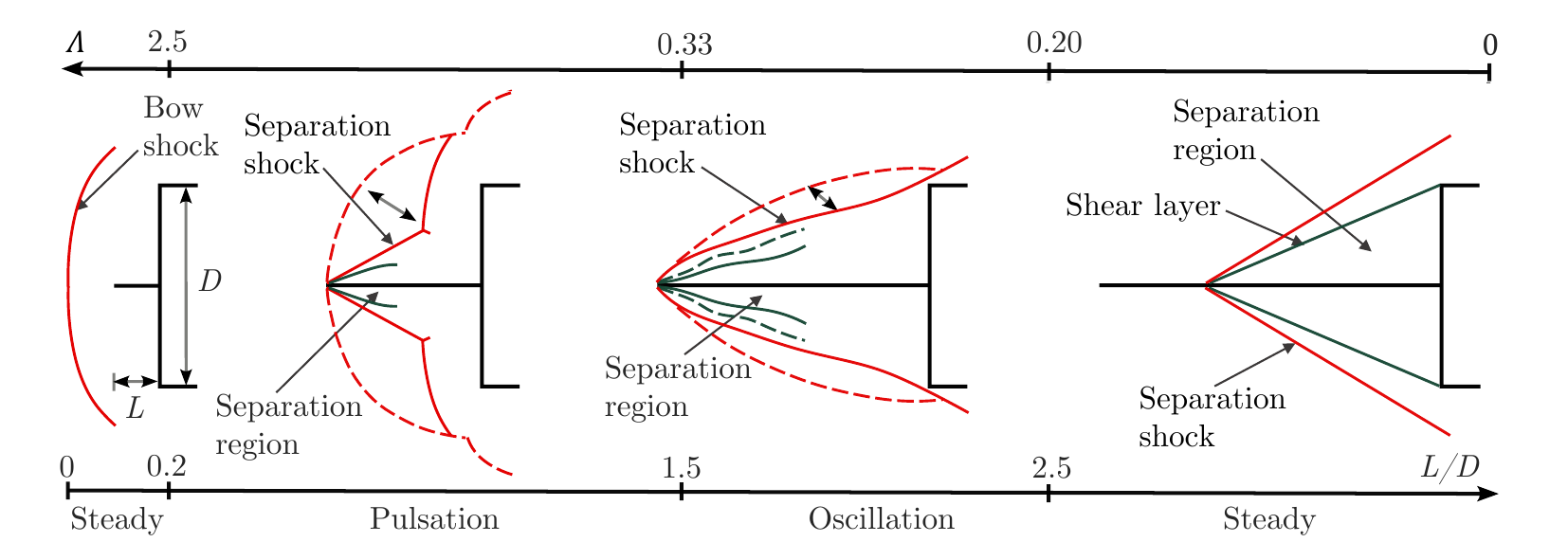}}
  \captionsetup{width=1\textwidth}
  \caption{An illustration of different flow states over a high-speed spiked cylinder ($\mathit{\Lambda}$ = $D/2L$).}
\label{fig:2}
\end{figure}

For given flow Reynolds and Mach numbers, $\mathit{\Lambda}$ solely governs the spike-cylinder flow behavior. A variation in $\mathit{\Lambda}$ can bring about extensive changes in the flow phenomena, leading to transitions between distinct regimes or states. Figure~\ref{fig:2} schematically illustrates the various flow states that occur across the $\mathit{\Lambda}$ parameter space, where $\mathit{\Lambda}$ = $D/2L$, with $D$ being the base cylinder diameter and $L$ the spike length.  For large values of $\mathit{\Lambda}$, the entire spike lies downstream of the bow shock generated by the cylinder, resulting in a nominally steady flow with subsonic conditions in the region between the shock and the cylinder face. A decrease in $\mathit{\Lambda}$ to the extent where the spike tip moves upstream of the bow shock wave brings about a sharp change in the flow with onset of large-amplitude shock wave unsteadiness, termed as “pulsations.” Flow pulsations are characterized by periodic shock-wave motion and separated flow along the entire spike length. A further decrease in $\mathit{\Lambda}$ leads to a transition in the flow state to a distinct, relatively smaller-scale unsteadiness in the separation shock wave, termed as “oscillations.” Flow oscillations are primarily driven by disturbances in the shear layer that forms over the separation region. Upon further decrease in $\mathit{\Lambda}$, the separation region remains localised in the vicinity of the spike base, nested by the corner, while a nominally steady shock system forms around the body. The $\mathit{\Lambda}$ boundaries between the different flow states shown in figure~\ref{fig:2} are nominal and exhibit a small degree of dependence on the flow Mach number, as suggested by existing experimental data from the literature \citep{Kenworthy1978,panaras1981pulsating,doi:10.2514/1.9035,panaras2009high,sasidharan2021large}. \par 

Lately, there is a growing interest in the double cone flow as a multi-parameter problem in the $\theta_1$–$\theta_2$–$\mathit{\Lambda}$ parameter space. \citet{sasidharan2021large} experimentally studied the double cone flow in the $\theta_1$-$\mathit{\Lambda}$ parameter space with $\theta_2$ fixed at $90^\circ$ (cone-cylinder model). \citet{Hornung_Gollan_Jacobs_2021} performed computations for the high-speed double cone flow while varying all three geometric parameters $\theta_1$, $\theta_2$, and $\mathit{\Lambda}$. Results from these studies show that even as the flow moves into a higher-dimensional geometric parameter space from the single-parameter spike-cylinder geometry, it only exhibits two distinct states of unsteadiness: pulsation and oscillation. These unsteady flow states are qualitatively the same as the pulsation and oscillation states reported in earlier literature on spike-cylinder flow.  These studies also show that the boundaries of the unsteady flow states in the $\theta_1$–$\theta_2$–$\mathit{\Lambda}$ parameter space are nonlinear. \par
Investigations into unsteady phenomena in the high-speed double cone flow have thus far been largely confined to obtaining a qualitative understanding based on physical observations. However, in recent years, there has been a growing emphasis on developing quantitative frameworks to better understand the spatiotemporal characteristics of flow unsteadiness. An advancement in this direction was made by \citet{Kumar_Sasidharan_Kumara_Duvvuri_2024}, who proposed an aeroacoustic model to predict the Strouhal number associated with the oscillatory mode of unsteadiness in double-cone configurations. In this model, the unsteadiness is attributed to a compressible shear layer that forms over the separation region and impinges on the aft-cone shoulder, thereby generating acoustic waves. These waves travel upstream through the subsonic separation bubble, forming a closed feedback loop, akin to the mechanism originally proposed by \citet{Rossiter1964Wind} for open-cavity flows. Based on a modified Rossiter-type formulation, their model showed good predictive accuracy for the Strouhal number (non-dimensional frequency) across the $\theta_1$–$\mathit{\Lambda}$ parameter space. \par

The broader motivation for the present effort is to develop an analytical model of pulsations in a high-speed double cone, and thereby obtain a deeper understanding of the underlying flow dynamics. As a starting point for a larger exercise, this paper focuses on the spike-cylinder, which is a particular configuration of the double cone. The current understanding of pulsations derived from available literature on high-speed spike-cylinder flow is qualitative in nature, and suggests pulsation is primarily an inviscid phenomenon \citep{Kenworthy1978,feszty2002utilising}. Studies by \citet{antonov1976nonsteady} and \citet{Kenworthy1978} hypothesized that flow pulsations arise due to periodic mass influx into the separated region, driven by interactions between the fore-shock and after-shock systems. This idea was further supported by \citet{panaras1981pulsating}. Subsequently, \citet{doi:10.2514/1.9034} extended and modified this hypothesis on the basis of their computational study. Although these efforts improved the physical understanding of the flow unsteadiness, they were not substantiated by a analytical framework or model. \par 

The present effort is aimed at obtaining a quantitative understanding of the dependence of spatiotemporal scales of pulsation on the governing geometric parameter in the spiked cylinder flow. The spike-cylinder configuration is particularly well suited for one-dimensional analysis and serves as a representative geometry for studying the fundamental features of pulsation-type unsteadiness. An analytical model that captures the underlying physics responsible for flow pulsations and predicts the associated Strouhal number is developed. In addition, a series of spike-cylinder experiments were performed at Mach 6, wherein the parameter $\mathit{\Lambda}$ was systematically varied while $\theta_1$ and $\theta_2$ were fixed at $0^\circ$ and $90^\circ$, respectively, to validate the model. The experimental set-up is described in \S \ref{Experimental setup}, and experimental results are presented in \S \ref{Experimental results}. Section \ref{Experimental results} also elaborates on the physical mechanisms that sustain flow pulsations. The analytical model for pulsation is developed in \S \ref{a model for pulsation}, and its predictions are compared with experimental measurements. Finally, a set of brief concluding remarks are presented in \S \ref{conclusions}.

\section{Experimental set-up}
\label{Experimental setup}
\begin{figure}
\centerline{\includegraphics[width=1\columnwidth]{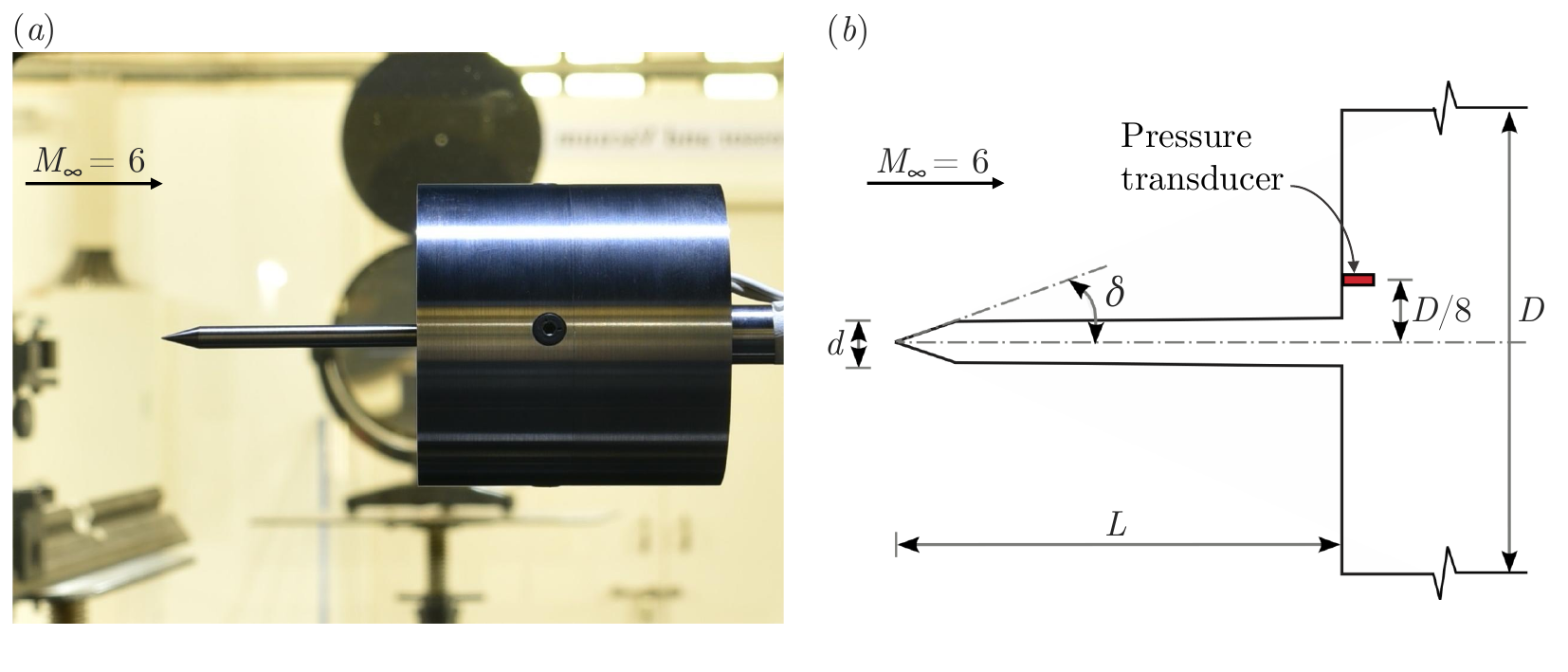}}
\captionsetup{width=1\textwidth}
\caption{$(a)$ Spike-cylinder model installed in the RNHWT test section. $(b)$ A line drawing of the spike-cylinder model (not to scale).}
\label{fig:3}
\end{figure}

Experiments were conducted in the Roddam Narasimha Hypersonic Wind Tunnel (RNHWT) at the Indian Institute of Science. The RNHWT is a 0.5-m-diameter enclosed free-jet facility (pressure-vacuum type) capable of producing freestream flows in the Mach number range 6–10, using dry air as the working fluid. A detailed description of the facility is provided by \citet{Thasu2024}. All experiments in the present study were carried out at a freestream Mach number of $M_{\infty} = 6$, with stagnation temperature and pressure set to $T_0 = 481$ K and $P_0 = 12.2$ bar, respectively. The corresponding freestream unit Reynolds number, calculated using the freestream density $\rho_{\infty}$, velocity $U_{\infty}$, and dynamic viscosity $\mu_{\infty}$, is $Re_{\infty} = 1\times 10^7$ m\textsuperscript{-1}. All spike-cylinder models used in this study are with fixed half-angles $\theta_1 = 0^\circ$ and $\theta_2 = 90^\circ$. The spike-cylinder model consists of a slender circular rod as the forebody and a right circular cylinder as the aftbody, aligned along a common axis. When $\theta_1$ is set to $0^\circ$ in a double-cone configuration, the fore-cone becomes a line coinciding with the aft-cone axis. For practical implementation, this line is replaced by a slender circular rod (\textit{i.e.}, the spike) with a finite diameter $d$, which is set as $0.075D$ in the present experiments. Figure~\ref{fig:3}(\emph{a}) shows a representative image of a spike-cylinder model installed in the RNHWT test section, while figure~\ref{fig:3}(\emph{b}) shows a line drawing. The spike tip is conical with a half-angle $\delta = 15^\circ$. A conical spike tip is chosen to ensure consistency with earlier studies \citep{wood1962hypersonic,antonov1976nonsteady,Kenworthy1978,doi:10.2514/1.9034,doi:10.1260/1475472054771367,sahoo2021shock}. The spike length was varied across experiments, and a total of six different configurations were studied with $\mathit{\Lambda}$ ($ =D/2L$) of 0.345, 0.385, 0.435, 0.500, 0.588, 0.714. The models were mounted with zero incidence and side-slip angles in all experiments. A miniature face-mounted pressure transducer (Endevco 8530C-100) was installed on the cylinder face at a radial distance of $D/8$ from the spike-cylinder axis, as indicated in figure~\ref{fig:3}(\emph{b}). This transducer has a sensing diameter of 2.4 mm and a resonance frequency of 500 kHz, with an effective bandwidth up to 100 kHz, which comfortably covers the frequency range of interest in this study.
Flow unsteadiness in all experiments was visualized using the technique of high-speed schlieren. A pulsed diode laser (Cavilux Smart, 640 nm wavelength, 10 ns pulse width) served as the light source and a high-speed camera (Phantom V1612) was used for imaging. Schlieren images were recorded at frame rates ranging from 40,000 to 60,000 frames per second. Across all experiments, the number of pulsation cycles captured in schlieren data range from 2,300 to 3,400, resulting in a temporal resolution of 20–37 frames per pulsation cycle. The resulting datasets offer sufficient spatiotemporal resolution and data length to enable detailed analysis.

\section{Experimental results}
\label{Experimental results}
\begin{figure}
  \centerline{\includegraphics[width=1\columnwidth]{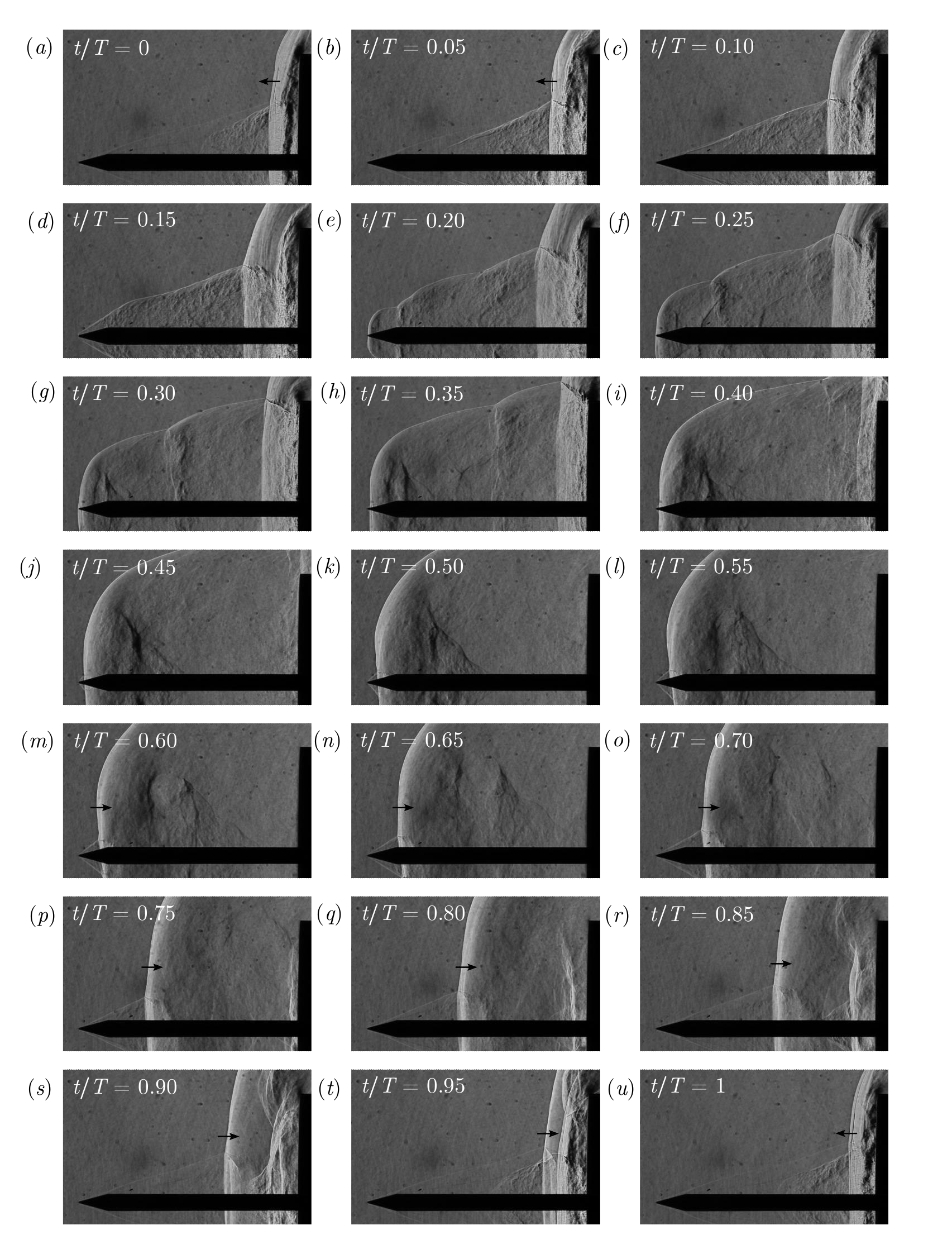}}
  \captionsetup{width=1\textwidth}
  \caption{A sequence of schlieren snapshots over one cycle of pulsation for $\mathit{\Lambda}$ = 0.5 and $M_\infty$ = 6. The cycle consists of three phases: axial inflation ($a$ to $d$);  radial inflation ($d$ to $i$); collapse ($i$ to $u$). Arrows marked over the bow shock indicate its instantaneous direction of motion.}  
\label{fig:4}
\end{figure}
\begin{figure}
  \centerline{\includegraphics[width=1\columnwidth]{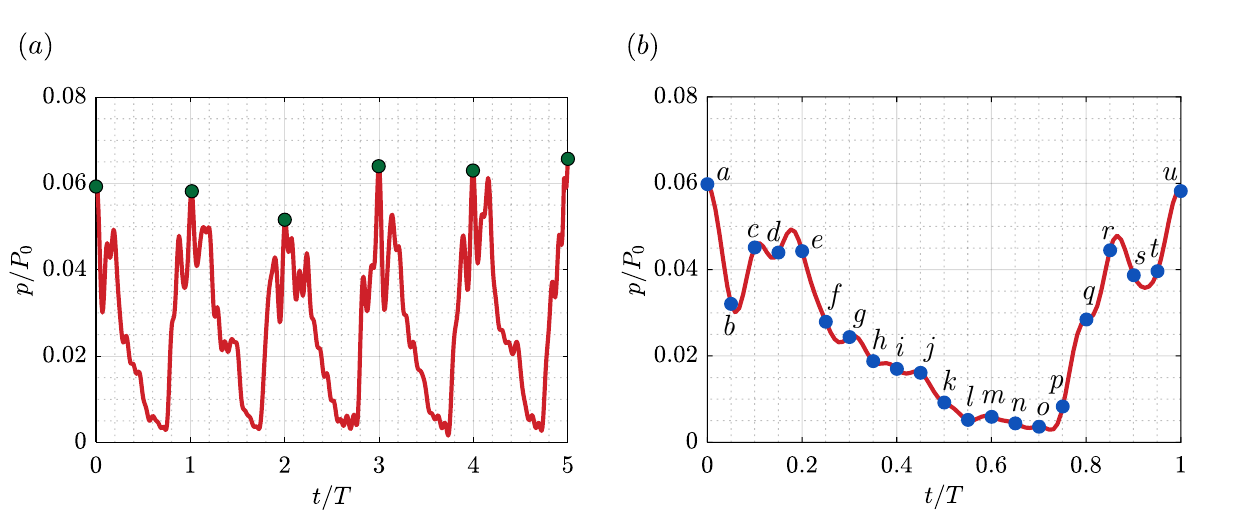}}
  \captionsetup{width=1\textwidth}
  \caption{$(a)$ Pressure signal recorded on the cylinder face at a radial distance of $D/8$ for $\mathit{\Lambda}$ = 0.5 over a representative period of five pulsation cycles. $(b)$ Pressure variation in one pulsation cycle, corresponding to the schlieren data shown in figure~\ref{fig:4}; the blue-colored markers and associated labels correspond to the time instants at which schlieren snapshots are shown in figure~\ref{fig:4}.
}
\label{fig:5}
\end{figure}
The large-scale periodic flow unsteadiness in pulsation can be best understood through a sequence of schlieren snapshots captured over a complete pulsation cycle. During pulsation, the shock system undergoes periodic motion with a time period $T$ and amplitudes comparable to the characteristic length scale $L$ of the spike. As a representative example, figure~\ref{fig:4} presents a sequence of schlieren images recorded at 21 distinct time instants over one pulsation cycle for $\mathit{\Lambda} = 0.5$. Given the axisymmetric nature of the flow, only the upper half of the flow field is shown in all schlieren images presented here. At $t/T = 0$ (figure~\ref{fig:4}$a$), the shock system consists of a bow shock generated by the cylindrical aft body and a conical separation shock originating from the separated flow region formed over the spike body. The conical separation shock progressively moves upstream (figures~\ref{fig:4}$a$ through~\ref{fig:4}$d$) and eventually reaches the spike tip (figure~\ref{fig:4}$d$). Subsequently, the conical shock propagates radially outward while a nearly normal shock forms at the spike tip, as seen in figures~\ref{fig:4}$(e)$ through \ref{fig:4}$(i)$. At around figure~\ref{fig:4}($j$), the combined shock structure of the normal shock and conical separation shock resembles a bow shock that fully envelops the spike–cylinder body. In the next phase of the cycle (figures~\ref{fig:4}$k$ through~\ref{fig:4}$u$), this bow shock collapses toward the cylinder face, thereby completing the pulsation cycle. \par

The wall pressure data for the above case ($\mathit{\Lambda} = 0.5$), measured on the cylinder face at a radial location of $D/8$ from the spike–cylinder axis, is presented in figure~\ref{fig:5}. The pressure $p$ is normalized by the freestream stagnation pressure $P_0$. In figure~\ref{fig:5}(\emph{a}) the local maxima in each pulsation cycle are highlighted by green-colored markers. Figure~\ref{fig:5}($b$) shows the temporal evolution of the normalized pressure over a single pulsation cycle, corresponding to the schlieren data presented in figure~\ref{fig:4}. The time instants corresponding to the schlieren frames (figures~\ref{fig:4}$a$ through \ref{fig:4}$u$) are marked in Figure~\ref{fig:5}($b$). \par

To obtain a detailed understanding of the flow during a pulsation cycle, the cycle is divided into three phases based on the motion of the shock system. These phases are labeled as “axial inflation,” “radial inflation,” and “collapse.” These phases are now described with reference to the pulsation cycle shown in figure~\ref{fig:4}. The starting point of the cycle, \textit{i.e.}, $t/T = 0$ (figure~\ref{fig:4}\textit{a}), is chosen to be the instant when the bow shock is closest to the cylinder face. The axial inflation phase, from figures~\ref{fig:4}\textit{(a)} to~\ref{fig:4}\textit{(d)}, corresponds to the upstream propagation of the conical separation shock until it reaches the spike tip. The radial inflation phase, spanning figures~\ref{fig:4}\textit{(d)} to~\ref{fig:4}\textit{(i)}, is characterized by the radial outward propagation of the separation shock away from the spike axis. Finally, the collapse phase, from figures~\ref{fig:4}\textit{(i)} to~\ref{fig:4}\textit{(u)}, involves the downstream motion of the bow shock as it retreats toward the cylinder face. The schlieren frames shown in figures~\ref{fig:4}\textit{(a)} and~\ref{fig:4}\textit{(u)} represent nearly identical stages of the cycle, thereby marking the completion of one full pulsation period. The characteristic flow features associated with each of these three phases are discussed in detail in the following subsections, and that forms the basis for the analytical model developed in \S~\ref{a model for pulsation}.

\subsection{Axial inflation} \label{axial inflation}
\begin{figure}
  \centerline{\includegraphics[width=1\columnwidth]{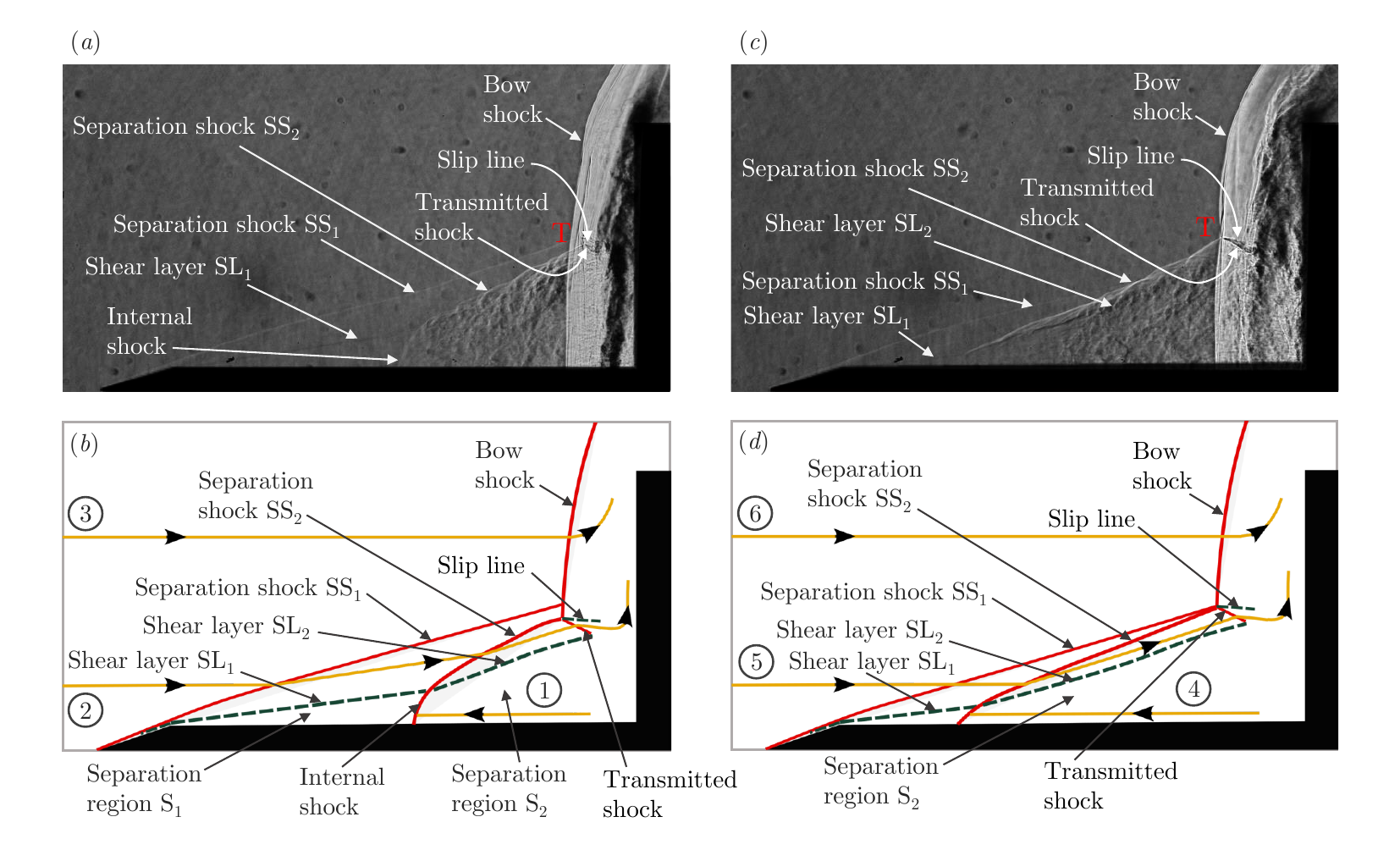}}
  \captionsetup{width=1\textwidth}
  \caption{($a$) A magnified view of figure~\ref{fig:4}($a$); ($b$) a schematic illustration of key flow features seen in ($a$). ($c$) A magnified view of figure~\ref{fig:4}($b$); ($d$) a schematic illustration of key flow features seen in ($c$).}
\label{fig:6}
\end{figure}
It is instructive to begin the discussion on pulsation with the axial inflation phase.  At the beginning of this phase ($t/T = 0$), the bow shock is at its closest position to the cylinder face, as seen in figure~\ref{fig:4}($a$). The pressure on the cylinder face reaches its maximum, indicated by point ($a$) in figure~\ref{fig:5}($b$). A magnified view of figure~\ref{fig:4}($a$) is presented in figure~\ref{fig:6}($a$), with key flow features marked, and a schematic illustration is shown in figure~\ref{fig:6}($b$). At this instant it is seen that the boundary layer is separated at the spike tip, forming a separation region downstream (labeled as $\mathrm{S_1}$) which is roughly conical in shape. This generates a conical shock (labeled $\mathrm{SS_1}$) at the spike tip. A shear layer, $\mathrm{SL_1}$, forms over the region $\mathrm{S_1}$. Further downstream, a second separation region, $\mathrm{S_2}$, is present along the spike. It generates a stronger conical shock, $\mathrm{SS_2}$, characterized by a larger shock angle than $\mathrm{SS_1}$. A second shear layer, $\mathrm{SL_2}$, forms over this region. These features are annotated in figure~\ref{fig:6}($b$). \par

During the axial inflation phase, the separation shock $\mathrm{SS_1}$ remains anchored at the spike tip, while the separation shock $\mathrm{SS_2}$ propagates toward the spike tip. This motion of $\mathrm{SS_2}$ is evident from figures~\ref{fig:4}($a$) through~\ref{fig:4}($d$). The lower portion of $\mathrm{SS_2}$, situated between the spike surface and shear layer $\mathrm{SL_1}$, shows a higher shock angle compared to the portion above $\mathrm{SL_1}$, due to different local upstream conditions. This lower segment of $\mathrm{SS_2}$, traveling upstream through the nearly stagnant gas in $\mathrm{S_1}$, is referred to as the ``internal shock'' and is highlighted in figure~\ref{fig:6}($a$). The upper portion, on the other hand, propagates through the conical flow downstream of $\mathrm{SS_1}$.
As the internal shock propagates upstream, the region $\mathrm{S_2}$ expands axially, progressing toward the spike tip. On the downstream side, the second separation shock, $\mathrm{SS_2}$, interacts with the bow shock, resulting in a transmitted shock. This interaction is clearly visible in figure~\ref{fig:6}($c$). A triple point `T' forms at the intersection of the bow, separation, and transmitted shocks, from which a slip line emerges due to velocity difference downstream of bow and transmitted shocks. The nature of the interaction between the separation and bow shocks falls under the type IV classification of \citet{edney1968anomalous}. The transmitted shock impinges on shear layer $\mathrm{SL_2}$, while a supersonic flow downstream of the transmitted shock, commonly known as Edney’s jet or the supersonic jet, moves toward the cylinder shoulder, where it expands around the shoulder corner. This jet trajectory is illustrated by streamline 5 in figure~\ref{fig:6}($d$). \par

It is important to note that during axial inflation, the freestream flow does not enter the separation region. Instead, it traverses the separation and transmitted shocks and is directed toward the cylinder shoulder, as indicated by streamline 5. This understanding is in contrast to earlier hypotheses by \citet{antonov1976nonsteady}, \citet{Kenworthy1978}, and \citet{panaras1981pulsating}, who proposed that axial inflation was driven by a continuous influx of freestream fluid through Edney’s jet. The present description is in agreement with \citet{doi:10.2514/1.9034}, who showed that the streamlines originating in the freestream are directed away from the separation region and toward the cylinder shoulder during the axial inflation phase. \par

The physical origin of axial inflation is suggested to be gas expansion. Specifically, at the end of the preceeding pulsation cycle, a high-pressure region is formed in front of the cylinder face. This gas expands into the low-pressure separation region $\mathrm{S_1}$, initiating the axial inflation. The resulting reverse flow creates the second separation region $\mathrm{S_2}$ and generates the separation shock $\mathrm{SS_2}$, which then propagates towards the spike tip. During this process, the pressure on the cylinder face shows a net decrease from points ($a$) to ($d$) in figure~\ref{fig:5}($b$). A second pressure peak is observed near point ($d$), which can plausibly be attributed to the impingement of Edney’s jet on the cylinder face \citep{doi:10.1260/1475472054771367}. During the axial inflation phase, the high-pressure gas primarily expands axially, although some radial motion of the separation shock $\mathrm{SS_2}$ is also observed. This radial movement is reflected in the slight increase in the height of the `T' from figure~\ref{fig:4}($a$) to~\ref{fig:4}($d$). The axial inflation phase concludes when the propagating shock $\mathrm{SS_2}$ reaches the spike tip and merges with $\mathrm{SS_1}$, forming a single separation shock (figure~\ref{fig:4}$d$), which will be referred to as the separation shock for the rest of the pulsation cycle. Early in the axial inflation phase, the bow shock undergoes a slight upstream shift, as observed in figures~\ref{fig:4}($a$) through~\ref{fig:4}($c$). In figure~\ref{fig:4}($c$), the shape and stand-off distance of the bow shock above the `T' closely match that of a forward-facing cylinder without a spike forebody (obtained from separate experiments). This stand-off distance is referred to as the regular stand-off distance, $x_b$ (at Mach 6, $x_b$ is close to $0.25D$). Once the bow shock attains a stand-off distance of $x_b$, it remains stationary, while the separation shock $\mathrm{SS_2}$ continues to propagate toward the tip for the rest of the axial inflation phase.

\subsection{Radial inflation} \label{radial inflation}
\begin{figure}
  \centerline{\includegraphics[width=1\columnwidth]{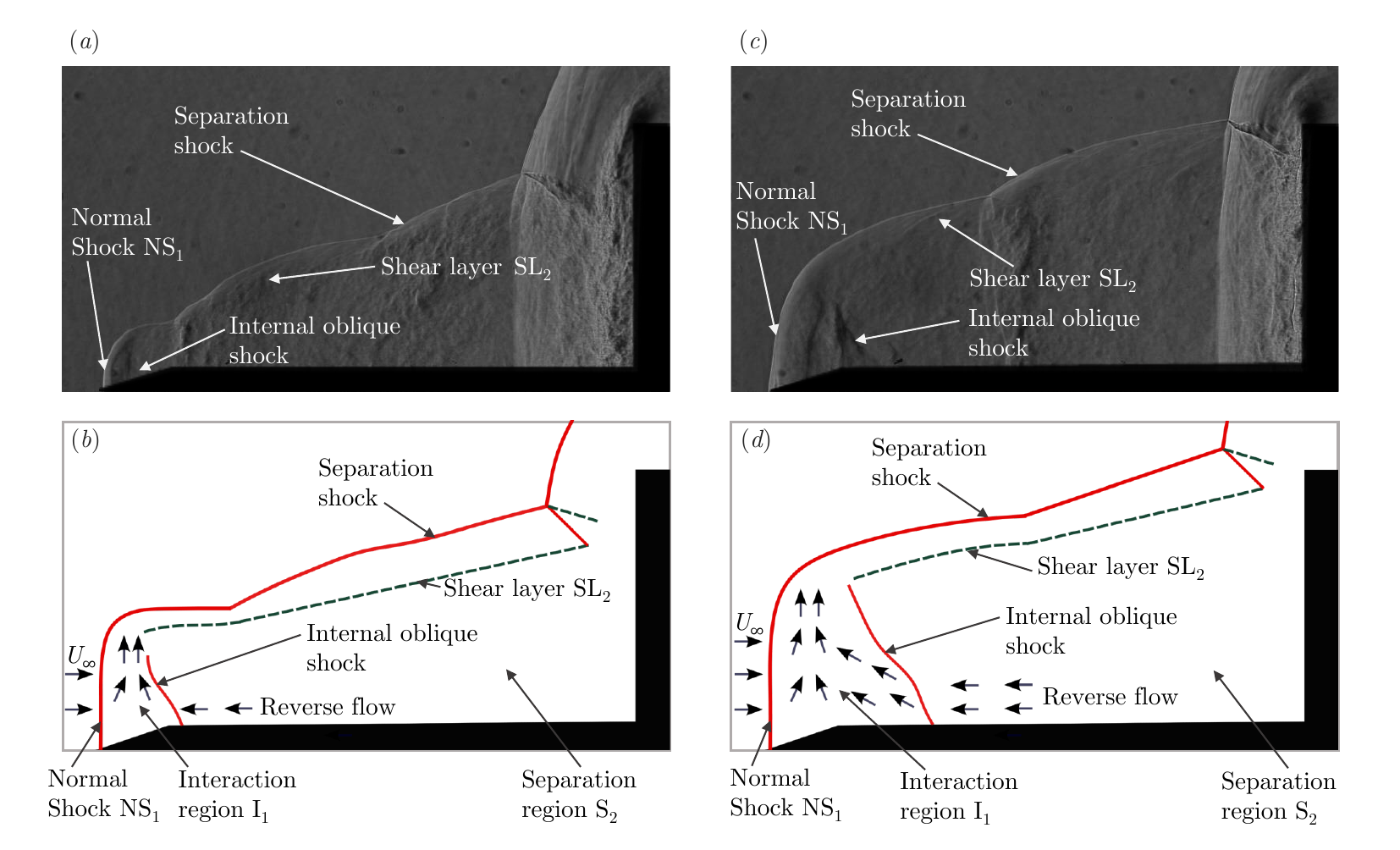}}
  \captionsetup{width=1\textwidth}
  \caption{($a$) A magnified view of figure~\ref{fig:4}($e$); ($b$) a schematic illustration of key flow features seen in ($a$). ($c$) A magnified view of figure~\ref{fig:4}($g$); ($d$) a schematic illustration of key flow features seen in ($c$).}
\label{fig:7}
\end{figure}
The radial inflation phase is characterized by the radial propagation of the conical separation shock away from the spike, as observed in figures~\ref{fig:4}($d$) through~\ref{fig:4}($i$). The high-pressure gas inside the separation region $\mathrm{S_2}$, which expands primarily in the axial direction during the axial inflation phase (discussed in \S~\ref{axial inflation}), now begins to expand radially. This radial expansion drives the outward movement of the separation shock and leads to a radial growth of region $\mathrm{S_2}$. During this phase, the pressure on the cylinder face experiences an overall decrease, as seen from points ($d$) to ($i$) in figure~\ref{fig:5}($b$). As the separation shock propagates outward, a nearly normal shock, labeled $\mathrm{NS_1}$, forms at the spike tip, and progressively increases in height throughout the radial inflation phase. The first appearance of $\mathrm{NS_1}$ is in figure~\ref{fig:4}($e$). A magnified view of this figure is presented in figure~\ref{fig:7}($a$), with a schematic illustration of the key flow features shown in figure~\ref{fig:7}($b$). As the freestream flow encounters $\mathrm{NS_1}$, the flow becomes subsonic, and subsequently the incoming gas collides with the strong reverse flow inside region $\mathrm{S_2}$ (this reverse flow is generated during the axial inflation phase). This interaction produces an internal oblique shock. The reverse flow passes through this oblique shock and then interacts with the subsonic flow downstream of $\mathrm{NS_1}$, resulting in a deflection in the radially outward direction. The region between $\mathrm{NS_1}$ and the internal oblique shock is identified as the interaction region $\mathrm{I_1}$, and is marked in figure~\ref{fig:7}($b$). The internal oblique shock grows in size and becomes increasingly prominent in the schlieren images from figures~\ref{fig:4}($e$) through~\ref{fig:4}($i$). Simultaneously, the size of the interaction region $\mathrm{I_1}$ and the height of $\mathrm{NS_1}$ continue to grow. This temporal growth is clearly evident in figure~\ref{fig:7}($c$), which shows a magnified view of figure~\ref{fig:4}($g$). \par

Downstream, the separation shock continues to interact with the bow shock at the triple point `T'. As the separation shock propagates radially outward, `T' shifts accordingly, a trend visible from figures~\ref{fig:4}($d$) to~\ref{fig:4}($i$). During this phase, the height of the normal shock $\mathrm{NS_1}$ keeps increasing, and by the time instant shown in figure~\ref{fig:4}($i$), it approximately attains the height of the cylinder, marking the end of the radial inflation phase. At this point, the combined structure of $\mathrm{NS_1}$ and the separation shock entirely envelops the spike-cylinder configuration (figure~\ref{fig:4}$i$). Following this, $\mathrm{NS_1}$ begins to move toward the cylinder, while the separation shock continues its outward radial propagation. This downstream motion of $\mathrm{NS_1}$ marks the beginning of the final phase in the pulsation cycle—the collapse phase.

\subsection{Collapse} \label{collapse}
At the onset of the collapse phase, the normal shock $\mathrm{NS_1}$ begins to move toward the cylinder face, while the separation shock continues its outward radial motion. During this period the combined structure of $\mathrm{NS_1}$ and the separation shock gradually transforms into a bow shock, as seen in figures~\ref{fig:4}($i$) to~\ref{fig:4}($j$). This bow shock, labeled $\mathrm{BS_1}$, continues to move toward the cylinder throughout the collapse phase, as seen in figures~\ref{fig:4}($j$) to~\ref{fig:4}($u$). At the beginning of this phase (figure~\ref{fig:4}$i$), the radial expansion of the separation region $\mathrm{S_2}$ (discussed earlier in \S~\ref{radial inflation}) is nearly complete, leaving behind a low-pressure zone within $\mathrm{S_2}$. The internal oblique shock, generated during the radial inflation phase, now propagates through this low-pressure region. The freestream flow encounters $\mathrm{NS_1}$, resulting in a pressure rise inside region $\mathrm{I_1}$, while the pressure inside $\mathrm{S_2}$ is expected to be relatively lower. Thus, a localized high-pressure region is formed near the spike tip, which is separated from the low-pressure zone $\mathrm{S_2}$ by the internal oblique shock. \par

\begin{figure}
  \centerline{\includegraphics[width=1\columnwidth]{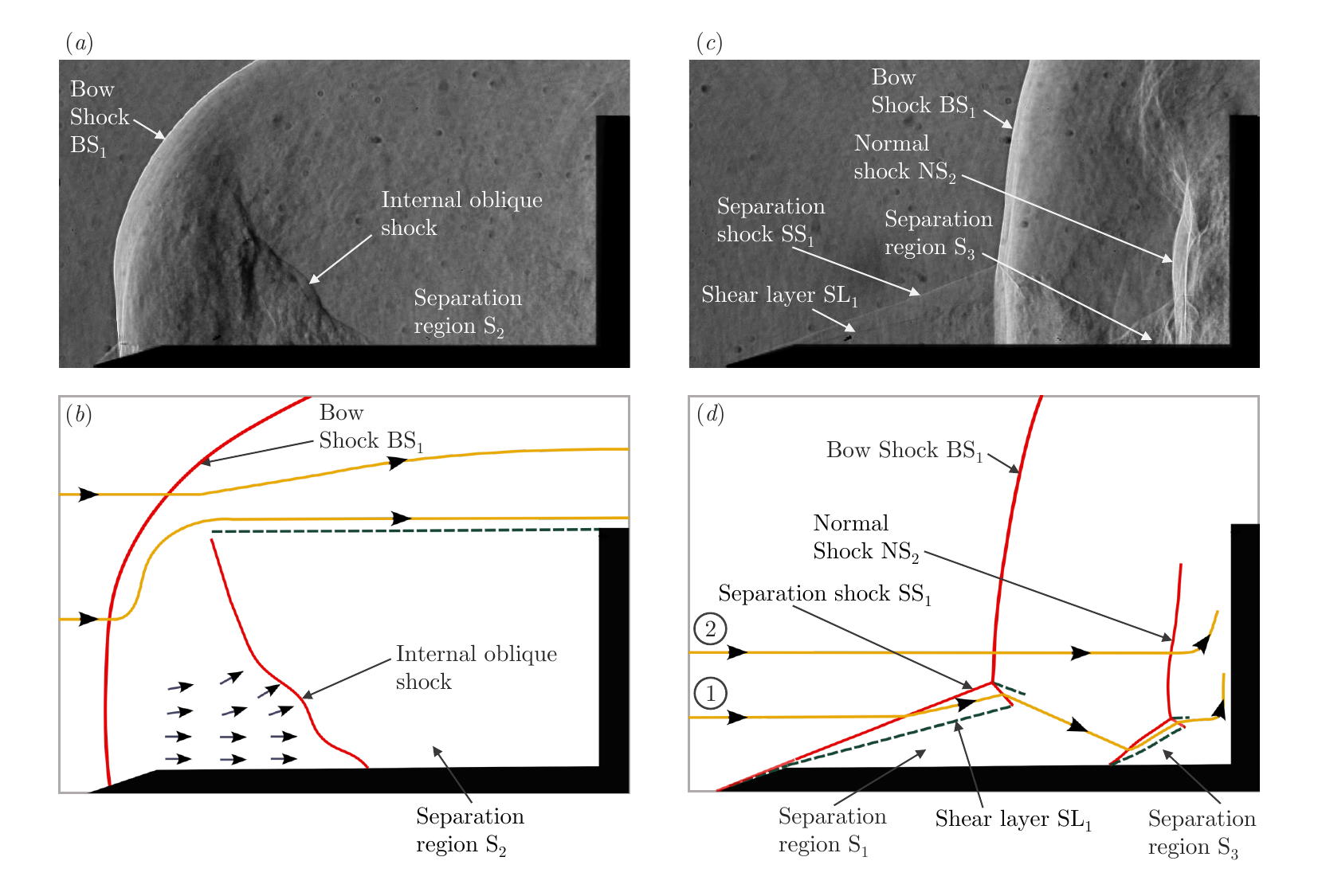}}
  \captionsetup{width=1\textwidth}
  \caption{($a$) A magnified view of figure~\ref{fig:4}($k$); ($b$) a schematic illustration of key flow features seen in ($a$). ($c$) A magnified view of figure~\ref{fig:4}($r$); ($d$) a schematic illustration of key flow features seen in ($c$).}
\label{fig:8}
\end{figure}

The internal oblique shock continues to propagate toward the cylinder until it impinges on the cylinder face, as observed from figures~\ref{fig:4}($i$) through~\ref{fig:4}($o$). During this interval the pressure on the cylinder face shows a slight decrease (figure~\ref{fig:5}$b$, points $i$ through $o$). While the internal oblique shock moves toward the cylinder, the bow shock also follows in the same direction. However, the internal oblique shock travels faster, causing the interaction region $\mathrm{I_1}$ to expand. Meanwhile, the reverse flow within $\mathrm{S_2}$, which is significant during the axial and radial inflation phases, progressively weakens. As a result, the flow field in region $\mathrm{I_1}$ undergoes a significant transformation. Where a colliding flow pattern previously existed (figure~\ref{fig:7}$b$), a predominantly downstream-directed flow now emerges, as shown in figure~\ref{fig:8}($b$). \par
The impingement of the internal oblique shock on the cylinder face generates a normal shock, labeled $\mathrm{NS_2}$, ahead of the cylinder. $\mathrm{NS_2}$ progressively strengthens and becomes increasingly prominent from figures~\ref{fig:4}($p$) to~\ref{fig:4}($s$). A close-up view of $\mathrm{NS_2}$ is shown in figure~\ref{fig:8}($c$). The supersonic flow from $\mathrm{I_1}$ encounters $\mathrm{NS_2}$, becomes subsonic, and experiences a pressure rise across the shock. It is worth noting here that supersonic conditions may develop downstream of a moving bow shock, whereas flow downstream of a steady bow shock is always subsonic. The pressure rise across $\mathrm{NS_2}$ results in an increase in pressure on the cylinder face, first observed at point ($p$) in figure~\ref{fig:5}($b$). At this stage, $\mathrm{NS_2}$ is first visible in the schlieren image of figure~\ref{fig:4}($p$). Downstream of $\mathrm{NS_2}$, the flow is directed toward the cylinder shoulder, where it undergoes expansion around the corner, as illustrated by streamline 2 in figure~\ref{fig:8}($d$). As $\mathrm{NS_2}$ strengthens for the rest of the collapse phase (figures~\ref{fig:4}$p$ to~\ref{fig:4}$u$), the pressure on the cylinder face rises, which is seen from points ($p$) to ($u$) in figure~\ref{fig:5}($b$). The bow shock $\mathrm{BS_1}$ continues to move toward the cylinder and eventually merges with $\mathrm{NS_2}$ in figure~\ref{fig:4}($u$), marking the end of the collapse phase, and the start of the axial inflation phase for the subsequent pulsation cycle. At this instant, the pressure on the cylinder face reaches its maximum (figure~\ref{fig:5}$b$).  During its downstream motion, the bow shock $\mathrm{BS_1}$ overshoots the regular standoff distance $x_b$ and merges with $\mathrm{NS_2}$ at a location closer to the cylinder. This final standoff distance is referred to as $x_n$. Present schlieren data shows that $x_n$ is consistently half of $x_b$, and this ratio remains invariant across all values of $\mathit{\Lambda}$. \par

During the collapse phase, while the bow shock $\mathrm{BS_1}$ moves toward the cylinder, the separation region $\mathrm{S_1}$ and a corresponding separation shock $\mathrm{SS_1}$ reappear. Early in this phase (figure~\ref{fig:4}$m$), a small separation region $\mathrm{S_1}$ forms near the spike tip. As the bow shock moves, this separation region enlarges, while its leading edge remains anchored at the spike tip; this is seen in figures~\ref{fig:4}($m$) through~\ref{fig:4}($u$). A shear layer $\mathrm{SL_1}$ forms over this region. In addition to $\mathrm{S_1}$, another separation region $\mathrm{S_3}$ appears near the spike base, induced by $\mathrm{NS_2}$ (figure~\ref{fig:8}$c$). Toward the end of the collapse phase, $\mathrm{S_3}$ merges with the region $\mathrm{S_1}$ (figure~\ref{fig:4}$t$), forming a large, low-pressure separation zone. This zone provides room for the high-pressure gas in front of the cylinder to expand into, triggering a strong flow reversal that initiates the axial inflation phase of the subsequent pulsation cycle.
\subsection{Quantitative characterization of flow pulsations} \label{Quantitative characterization of flow pulsations}
To quantify the unsteady behavior of the pulsation phenomenon, the pulsation Strouhal number (non-dimensional frequency) is extracted from the experimental data, and its dependence on the geometric parameter $\mathit{\Lambda}$ is examined. The methodology for estimating Strouhal number is presented here. And subsequently, the maximum pressure on the cylinder surface attained during the pulsation cycle is studied.
\subsubsection{Strouhal number estimation}
\begin{figure}
  \centerline{\includegraphics[width=0.9\columnwidth]{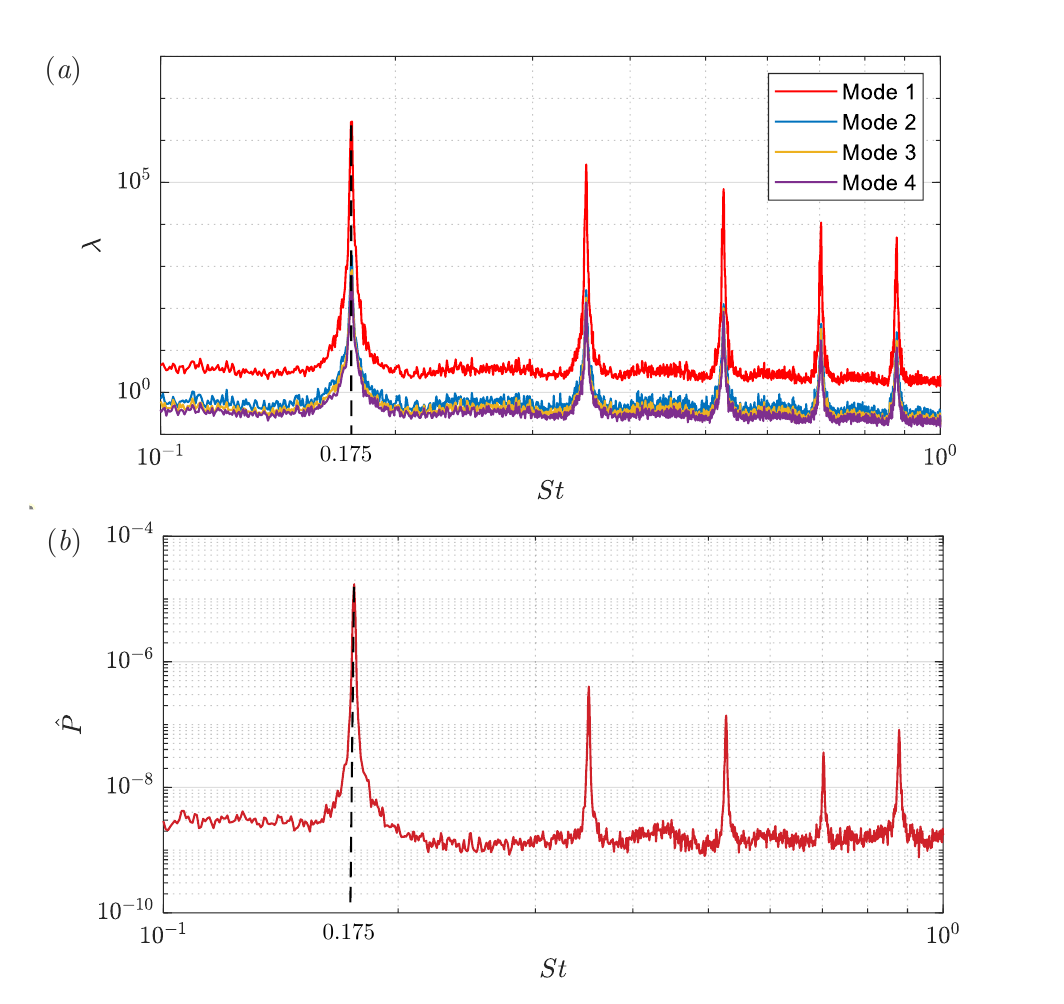}}
  \captionsetup{width=1\textwidth}
  \caption{($a$) SPOD modal energy spectra for $\mathit{\Lambda}$ = 0.5; ($b$) Power spectral density ($\hat{P}$) of non-dimensionalized pressure fluctuations $p'/P_0$ for $\mathit{\Lambda}$ = 0.5.
}
\label{fig:9}
\end{figure}

Qualitative observations from all schlieren datasets acquired in this study reveal that the unsteady motion of the shock system exhibits a single dominant time scale for any given $\mathit{\Lambda}$. This is consistent with observations reported in literature on pulsation in high-speed spike-cylinder flows. To quantitatively extract this global time scale, we apply spectral proper orthogonal decomposition (SPOD) to the high-speed schlieren data \citep{thasu2025universal,thasu2022strouhal}. In this technique the SPOD modes are ranked according to their eigenvalues $\lambda$, with the leading mode representing the dominant coherent structure in the flow \citep{,towne2018spectral}. Figure~\ref{fig:9}($a$) presents the SPOD results for the $\mathit{\Lambda} = 0.5$ case as a representative example. The analysis yields a Strouhal number $St = 0.175$, where $St$ is defined as:
\begin{equation}
St = \frac{fD}{U_{\infty}}.
\end{equation}
Here $f$ is the pulsation frequency, $D$ is the cylinder diameter, and $U_{\infty}$ is the free-stream velocity. As a confirmation of the $St$ obtained from SPOD, figure~\ref{fig:9}($b$) shows the power spectral density (PSD) of surface pressure fluctuations for the same case ($\mathit{\Lambda} = 0.5$). The peak in the PSD at $St = 0.175$ clearly confirms that the Strouhal number estimate obtained from SPOD analysis is accurate. SPOD was performed for data from all six $\mathit{\Lambda}$ configurations to estimate $St$, and these results will be presented in \S~\ref{a model for pulsation}. 
\subsubsection{Maximum pressure estimation}
\begin{figure}
  \centerline{\includegraphics[width=0.9\columnwidth]{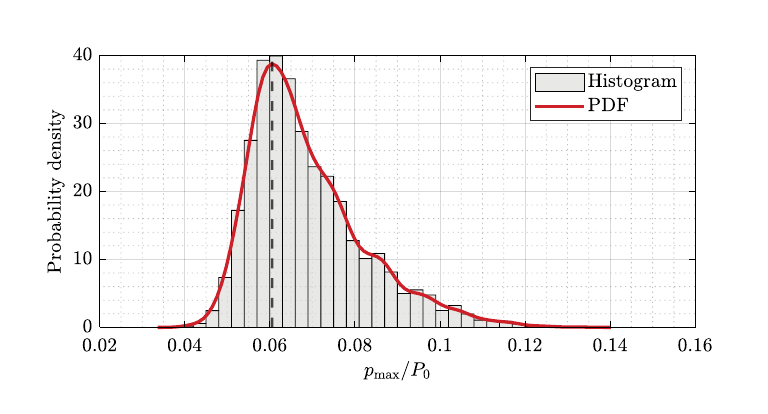}}
 \captionsetup{width=1\textwidth}
  \caption{ Probability density function (PDF) for maximum pressure $p_{\text{max}}/P_0$ over 2000 pulsation cycles for $\mathit{\Lambda} = 0.5$. The vertical dashed line at 0.0605 marks the most probable value of $p_{\text{max}}/P_0$.
}
\label{fig:10}
\end{figure}
\begin{figure}
  \centerline{\includegraphics[width=0.9\columnwidth]{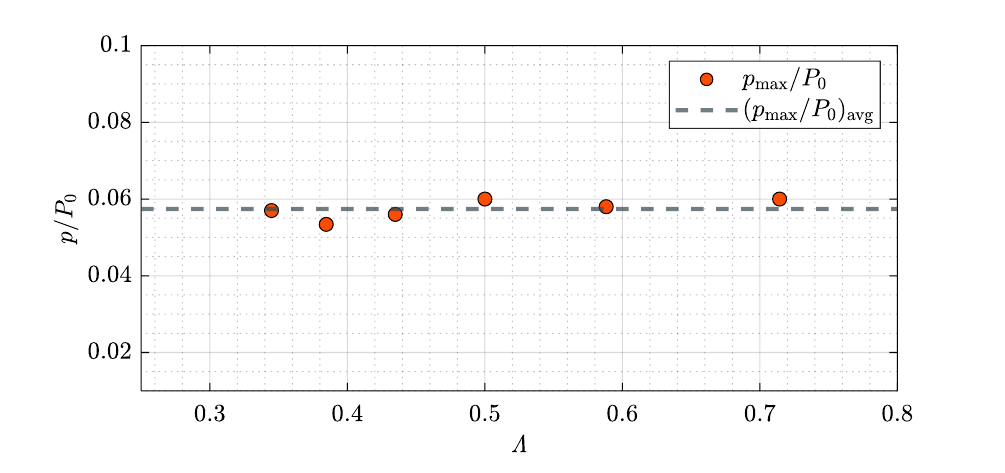}}
  \captionsetup{width=1\textwidth}
  \caption{Maximum pressure $p_{\text{max}}/P_0$ during pulsation for different $\mathit{\Lambda}$ configurations. Red markers indicate experimental measurements, and the dashed line represents the average value across the six experimental data points.
}
\label{fig:11}
\end{figure}

The maximum instantaneous pressure, $p_{\text{max}}$, from each pulsation cycle is recorded over 2000 cycles, and the probability density function (PDF) of the normalized maximum pressure $p_{\text{max}}/P_0$ is plotted in figure~\ref{fig:10} for $\mathit{\Lambda} = 0.5$. As a representative estimate of the maximum pressure, the most probable value from the PDF in figure~\ref{fig:10} is taken, which is $p_{\text{max}}/P_0 = 0.0605$ for $\mathit{\Lambda} = 0.5$. The same methodology was applied to all $\mathit{\Lambda}$ configurations, and the results are compiled in figure~\ref{fig:11}. It shows that $p_{\text{max}}/P_0$ remains nearly constant across different geometries, exhibiting no systematic dependence on $\mathit{\Lambda}$. Therefore, the maximum pressure attained during pulsation cycles is considered approximately invariant with respect to geometry over the range of $\mathit{\Lambda}$ studied here. By averaging over all six $\mathit{\Lambda}$ cases, a mean value of $p_{\text{max}}/P_0 = 0.0574$ is obtained.\par

It is noteworthy that the observed peak pressure $p_{\text{max}}$ is substantially higher than the downstream pressure associated with a steady bow shock for a forward-facing cylinder without a spike. For such a configuration, assuming a normal shock in the vicinity of the model centerline, the post-shock pressure on the cylinder axis, $p_b$, is given by the normal shock relation:
\begin{equation}
p_b = p_{\infty}\left[ 1 + \frac{2\gamma}{\gamma+1} \left( M_{\infty}^2 - 1 \right)\right],
\end{equation}
where $p_\infty$ and $M_\infty$ are the freestream pressure and Mach number, and $\gamma$ is the ratio of specific heats for air. For the present conditions, this yields $p_b = 0.0292P_0$, which implies that the peak pressure during pulsation is approximately twice as high as the steady-state bow shock pressure for the same cylinder without a spike.

\section{A model for pulsation}
\label{a model for pulsation}
Based on the physical understanding of the pulsation cycle derived from the previous section, an analytical model of the pulsation phenomenon is now developed. The overarching feature of flow pulsation is alternating cycles of gas compression and expansion that occur over the spike. At the end of the collapse phase a high-pressure region forms in front (upstream) of the cylinder face. As the subsequent pulsation cycle commences, this high-pressure gas initially expands axially (\emph{i.e.}, axial inflation phase) and later expands radially (\emph{i.e.}, radial inflation phase). This expansion pushes the shock system toward the spike tip. Once the expansion is complete, a low-pressure region develops in front of the cylinder, causing the shock system to collapse back toward the cylinder. This collapse regenerates the high-pressure region ahead of the cylinder, thereby completing the pulsation cycle. Building on this picture, each of the three phases--axial inflation, radial inflation, collapse--are modeled separately. The time duration of each phase predicted by the models are compared with experimental measurements. The total time period of the pulsation cycle is then simply obtained by summing the durations of the three individual phases. Finally, the non-dimensional pulsation frequency (Strouhal number) predicted by the model is compared against the experimental data.

\subsection{A model for axial inflation} \label{axial inflation model}

\begin{figure}
  \centerline{\includegraphics[width=0.9\columnwidth]{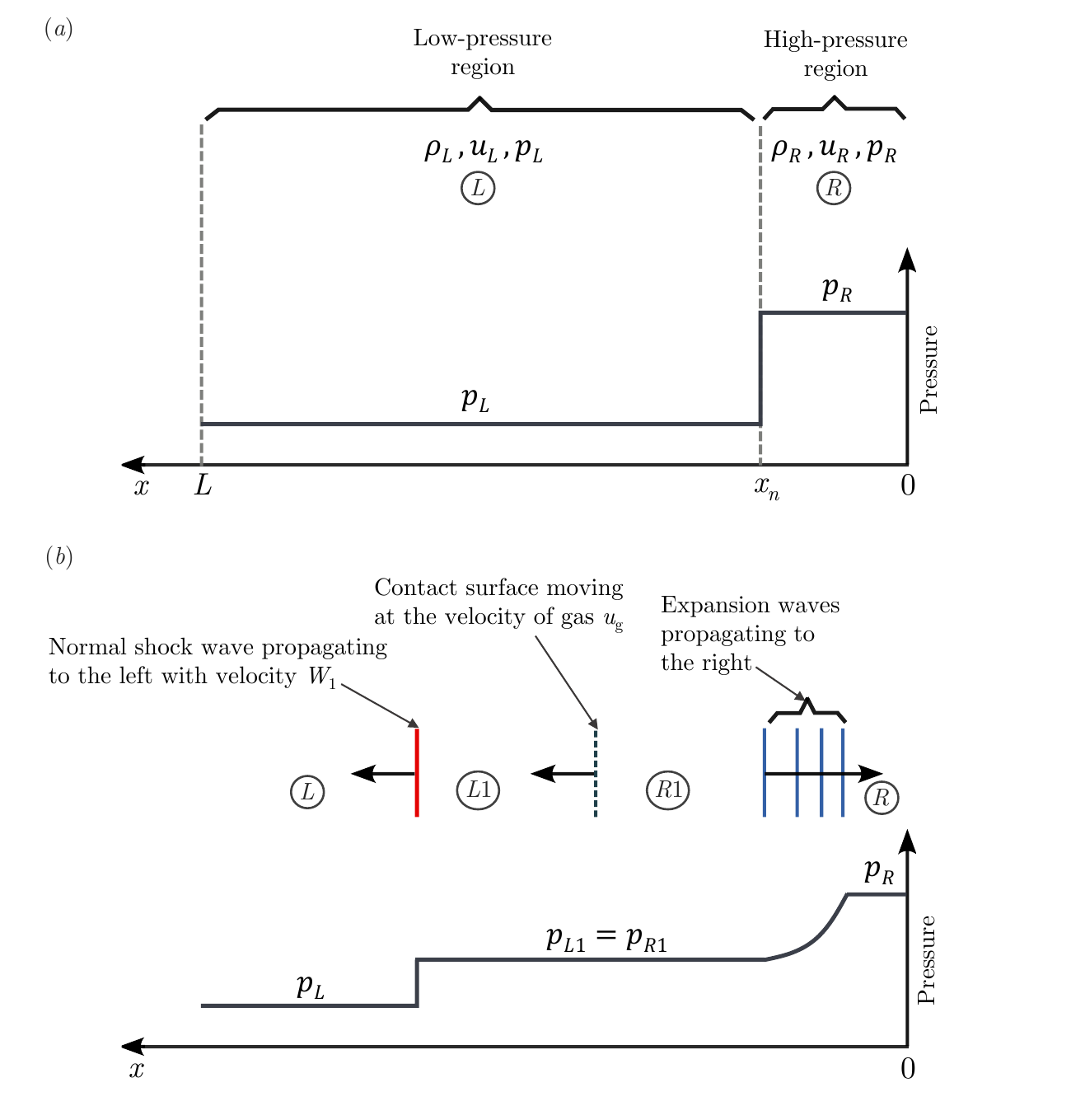}}
  \captionsetup{width=1\textwidth}
  \caption{($a$) Initial conditions for the axial inflation phase; ($b$) idealized one-dimensional flow in the axial inflation phase.}
\label{fig:13}
\end{figure}

The axial inflation phase is initiated by the expansion of high-pressure gas that forms in front the cylinder face at the end of the collapse phase of the preceding pulsation cycle. This gas expands into the low-pressure separation region $\mathrm{S_1}$. As the high-pressure gas flows into this region, it drives a separation shock ($\mathrm{SS_2}$) toward the spike tip. The end of the axial inflation phase is marked by $\mathrm{SS_2}$ reaching the spike tip. Hence, the duration of axial inflation is set by the time taken by $\mathrm{SS_2}$ to traverse the length of the spike in the upstream direction.
\begin{figure}
  \centerline{\includegraphics[width=0.7\columnwidth]{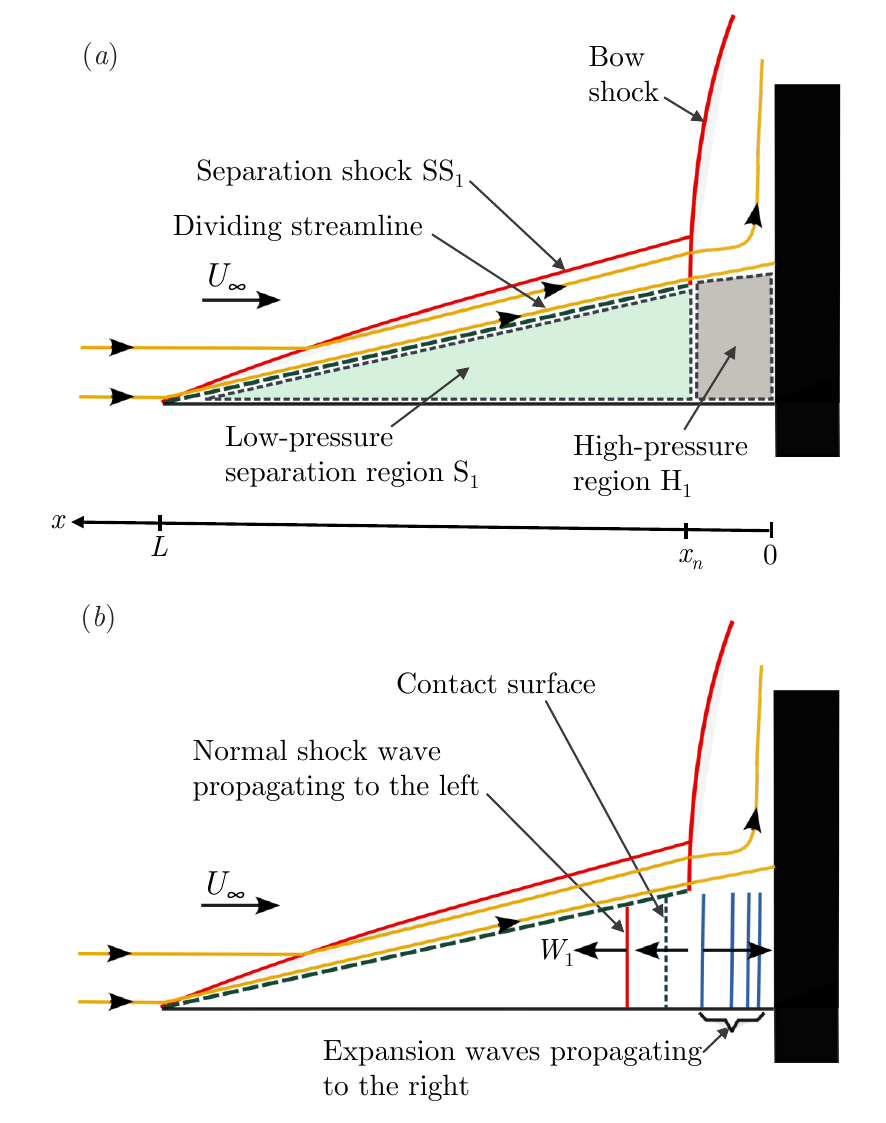}}
  \captionsetup{width=1\textwidth}
  \caption{A schematic representation of the axial inflation model. (a) shows the initial condition for the Riemann problem and (b) shows a representative time instant during the evolution of the solution to the Riemann problem. 
}
\label{fig:12}
\end{figure}
This is treated as a Riemann problem involving a high-pressure region $\mathrm{H_1}$ and a low-pressure separation region $\mathrm{S_1}$, separated by an initial discontinuity at the axial location $x = x_n$, where $x_n$ denotes the minimum bow shock stand-off distance (at the beginning of axial inflation). This setup is schematically illustrated in figure~\ref{fig:12}($a$), where the high-pressure region $\mathrm{H_1}$ and the low-pressure separation region $\mathrm{S_1}$ are marked. Note that the flow from the freestream moves toward the cylinder shoulder and does not enter the high-pressure region $\mathrm{H_1}$ during the axial inflation phase (as discussed in \S~\ref{axial inflation}). Therefore, a dividing streamline can be defined between the outer flow and the high-pressure region $\mathrm{H_1}$; this is indicated in figure \ref{fig:12}($a$). At the start of the axial inflation phase the gas in the $\mathrm{H_1}$ and $\mathrm{S_1}$ regions is assumed to be stagnant. Although a recirculating flow is present inside the separation region $\mathrm{S_1}$, it is neglected for simplicity since the characteristic velocity scale of the recirculating gas is expected to be much lower than the other characteristic velocity scales relevant to the problem. Uniform flow properties are assumed inside the regions $\mathrm{H_1}$ and $\mathrm{S_1}$, and the expansion process is assumed to be only in the axial direction. Consequently, the flow is treated to be one-dimensional in the regions $\mathrm{H_1}$ and $\mathrm{S_1}$, where all flow properties are only dependent on $x$ (and time $t$), which is the axis of symmetry of the spike-cylinder configuration. The motion of the shock wave is largely an inviscid phenomena, where the fluid viscosity is not expected to have a first-order effect. Therefore, the flow in regions $\mathrm{H_1}$ and $\mathrm{S_1}$ is treated by the one-dimensional inviscid conservation equations, which are written as  
\begin{equation}
\frac{\partial}{\partial t}
\begin{bmatrix}
\rho \\ \rho u \\ E
\end{bmatrix}
+
\frac{\partial}{\partial x}
\begin{bmatrix}
\rho u \\ \rho u^2 + p \\ (E + p)u
\end{bmatrix}
= 0,
\label{eq:invicid_cons}
\end{equation}
where $\rho$ is the density, $u$ is the velocity in the $x$-direction, $p$ is the pressure, and $t$ is time. $E$ is the total energy per unit volume, given by
\begin{equation}
E = \frac{p}{\gamma - 1} + \frac{1}{2} \rho u^2,
\end{equation}
where $\gamma$ is the specific heat ratio for air. The initial condition is written as
\begin{equation}
\begin{bmatrix}
\rho \\ u \\ p 
\end{bmatrix} (x, 0) =
\begin{cases}
\begin{bmatrix}
\rho_R \\ u_R \\ p_R
\end{bmatrix}, & x < x_n, \\\\ 
\begin{bmatrix}
\rho_L \\ u_L \\ p_L
\end{bmatrix}, & x \geq x_n.
\end{cases}
\label{eq:inf_initial}
\end{equation} 
Here, subscripts $L$ and $R$ denote the left and right states, corresponding to the separation region $\mathrm{S_1}$ and the high-pressure region $\mathrm{H_1}$, respectively. Initially, the low-pressure separation region and the high-pressure region are separated by a discontinuity located at $x = x_n$. This one-dimensional setup is schematically illustrated in figure~\ref{fig:13}($a$). The solution to this Riemann problem consists of a normal shock wave propagating left into the low-pressure region, while expansion waves propagate right into the high-pressure region \citep{Liepmann2002Elements}, as illustrated in figure~\ref{fig:13}($b$).
A contact discontinuity (contact surface) that moves left with velocity $u_g$ is a part of the solution. The region between the contact surface and the normal shock is labeled $L1$, while the region between the tail of the expansion fan and the contact surface is labeled $R1$. As the normal shock wave propagates left with velocity $W_1$, it induces gas motion with velocity $u_g$ in the region $L1$. Across the contact surface, both pressure and velocity remain continuous, \textit{i.e.}, $p_{L1} = p_{R1}$ and $u_{L1} = u_{R1} = u_g$. The expansion waves propagate into the high-pressure region $R$, reducing the pressure therein. The shock velocity $W_1$ and the associated flow properties are fully determined by the initial conditions in regions $L$ and $R$. In this analysis, air is assumed to behave as a calorically perfect gas. The normal shock relations are used to obtain the normal shock velocity $W_1$ as a function of the pressure ratio ($p_{L1}/p_{L}$) across the shock
\begin{equation}
W_1 = a_L\sqrt{\frac{\gamma + 1}{2\gamma}\left(\frac{p_{L1}}{p_{L}} - 1\right) + 1},
\label{eq:shock_speed}
\end{equation}
where $a_L$ = $\sqrt{\gamma p_L/\rho_L}$ is the  speed of sound in the region $L$. The velocity $u_g$ is given by
\begin{equation}
u_g = \frac{a_L}{\gamma} \left( \frac{p_{L1}}{p_L} - 1 \right)
\left( 
\frac{\frac{2\gamma}{\gamma + 1}}
{\frac{p_{L1}}{p_L} + \frac{\gamma - 1}{\gamma + 1}}
\right)^{1/2}.
\label{eq:particle_speed}
\end{equation}
The shock strength ($p_{L1}/p_{L}$) can be written as an implicit function of the initial pressure ratio ($p_{R}/p_{L}$) between the right and left states \citep{Liepmann2002Elements} 
\begin{equation}
\frac{p_R}{p_L} = \frac{p_{L1}}{p_L} \left\{ 1 - 
\frac{(\gamma - 1)(a_L / a_R)(p_{L1} / p_L - 1)}
{\sqrt{2 \gamma \left[ 2 \gamma + (\gamma + 1)(p_{L1} / p_L - 1) \right]}}
\right\}^{-\frac{2 \gamma}{\gamma - 1}}.
\label{eq:shock_strength}
\end{equation}
Using equation~\ref{eq:shock_strength} in equation~\ref{eq:shock_speed}, the shock velocity $W_1$ can be readily calculated. Assuming that the separation shock $\mathrm{SS_2}$ propagates with a constant velocity $W_1$ (see figure~\ref{fig:12}$b$), the time duration of the axial inflation phase can be estimated as
\begin{equation}
T_a = \frac{L-x_n}{W_1}. 
\label{eq:axial inflation_time}
\end{equation}

\subsubsection{Initial conditions for the axial inflation phase}

For the above formulation, suitable initial conditions in the left-side low-pressure region and the right-side high-pressure region must be defined. In the right-side high-pressure region $ R $, the initial pressure $p_R$ is taken as uniform and equal to the maximum pressure observed during the pulsation cycle, which was found to be approximately constant at $ 0.0574 P_0 $ for all $\mathit{\Lambda}$ cases (as discussed in \S  \ref{Experimental results}). The flow is assumed to be stagnant within this region, \emph{i.e.}, $u_R = 0$, and therefore the temperature is assumed to be equal to the free-stream stagnation temperature $T_0$. The initial conditions in region $R$ can then be expressed as
\begin{equation}
\left.
\begin{aligned}
\rho_R &= \frac{0.0574P_0}{R_{\text{gas}}T_0}, \\
u_R &= 0, \\\\
p_R &= 0.0574P_0,
\end{aligned}
\quad \right\} 
\label{eq:inf_right}
\end{equation}
where $R_{\text{gas}}$ is the universal gas constant. \par
\par
\begin{figure}
  \centerline{\includegraphics[width=0.7\columnwidth]{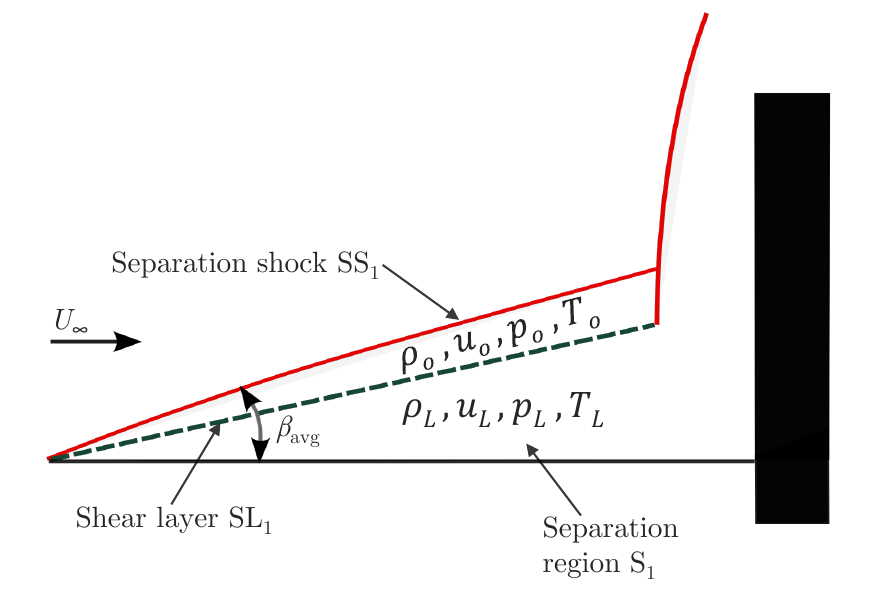}}
  \caption{Flow properties at the beginning of the axial inflation phase.}
\label{fig:14}
\end{figure}
The initial conditions for the left-side region are taken to be the flow properties inside the separation region $\mathrm{S_1}$ at the onset of the axial inflation phase. To determine these properties, the post-shock flow conditions downstream of the separation shock $\mathrm{SS_1}$ (see figure~\ref{fig:14}), denoted with subscript $o$, are estimated by solving the Taylor-Maccoll equation \citep{anderson2003modern} with $M_{\infty}=6$ and the shock angle $\beta$ obtained from schlieren data. Table~\ref{tab:1} lists the $\beta$ values for different $ \mathit{\Lambda} $ configurations, where it is seen that with increasing $\mathit{\Lambda}$ there is a slight increase in $\beta$. To simplify the analysis, an average shock angle of $\beta_{\text{avg}} = 16.25^\circ$ is used for all $\mathit{\Lambda}$ configurations. Once the conditions downstream to $\mathrm{SS_1}$ are known, the flow properties inside the separation region $\mathrm{S_1}$ (left side region $L$) can be estimated. Inside $\mathrm{S_1}$ the flow is assumed to be uniform and stagnant. Across the shear layer $\mathrm{SL_1}$, pressure continuity is assumed, \textit{i.e.}, $p_L = p_o$.
The velocity in region  $L$ is zero, \textit{i.e.}, $u_L = 0$. The temperature in this region is taken as the recovery temperature to account for viscous effects in the shear layer \citep{Anderson2017Fundamentals} 
\begin{table}
  \begin{center}
\def~{\hphantom{0}}
  \begin{tabular}{cc}
      $\mathit{\Lambda}$  & $\beta$ \\[6pt]
       0.345    & 13.7 \\
       0.385  & 15.2 \\
       0.435  & 16.5 \\
       0.500   & 16.8 \\
       0.588 & 17 \\
       0.714 & 18.3 \\
  \end{tabular}
  \caption{Shock angle $\beta$ at the start of the axial inflation phase, obtained from schlieren data.}
  \label{tab:1}
  \end{center}
\end{table}
\begin{equation}
T_L = \frac{1 + \sqrt{Pr} \left( \frac{\gamma - 1}{2} \right) M_o^2}{1 + \left( \frac{\gamma - 1}{2} \right) M_o^2}\, T_0,
\end{equation}
where $M_o$ is Mach number above the shear layer $\mathrm{SL_1}$ and $Pr$ is the Prandtl number for air (taken to be $0.71$). With that, the initial conditions for the left side region $L$ can be written as
\begin{equation}
\left.
\begin{aligned}
\rho_L &= \frac{p_L}{R_{\text{gas}}T_L} = \left(\frac{p_o}{R_{\text{gas}}}\frac{1 + \left( \frac{\gamma - 1}{2} \right) M_o^2}{\left(1 + \sqrt{Pr} \left( \frac{\gamma - 1}{2} \right) M_o^2\right)T_o}\right), \\
u_L &= 0, \\\\
p_L &= p_o.
\end{aligned}
\quad \right\} 
\end{equation}
Using these initial conditions, the normal shock strength (equation~\ref{eq:shock_strength}) is evaluated, which in turn provides the shock velocity $W_1$ through equation~\ref{eq:shock_speed}. Finally, the axial inflation time $T_a$ is estimated using equation~\ref{eq:axial inflation_time}.

\subsubsection{Model prediction for the axial inflation phase}
\begin{figure}
  \centerline{\includegraphics[width=0.9\columnwidth]{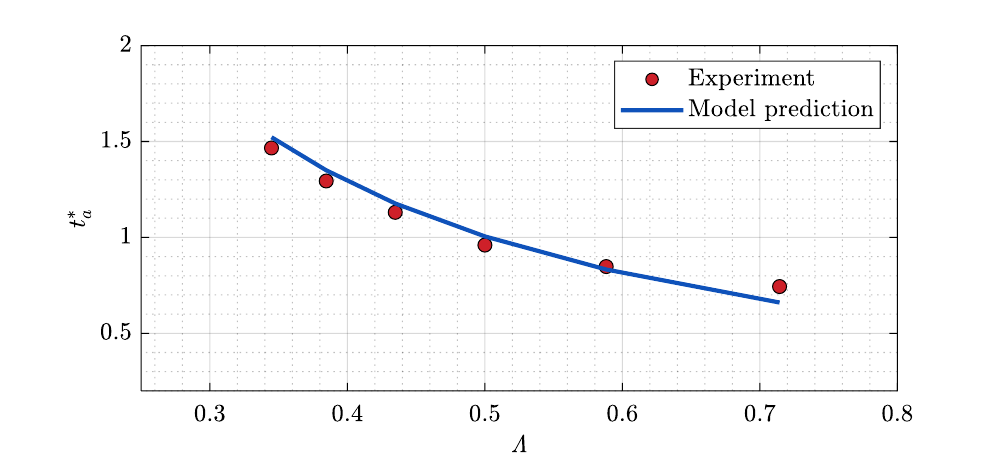}}
  \captionsetup{width=1\textwidth}
  \caption{A comparison of predicted non-dimensional axial inflation time $t^*_a \left(= \frac{T_aU_{\infty}}{D}\right)$ with experiments.
}
\label{fig:15}
\end{figure}

Using the above formulation the time duration of the axial inflation phase is estimated for the six different values of $\mathit{\Lambda}$ that were studied experimentally. Figure~\ref{fig:15} presents a comparison of the non-dimensional axial inflation time, defined as $t^*_a = \frac{T_a U_{\infty}}{D}$, between experimental measurements and model predictions. The model clearly captures the trend of $t^*_a$ decreasing with $\mathit{\Lambda}$, and also shows good quantitative agreement with the experimental measurements. 

\subsubsection{Flow properties inside the separation region $\mathrm{S_2}$}
\begin{figure}
  \centerline{\includegraphics[width=1\columnwidth]{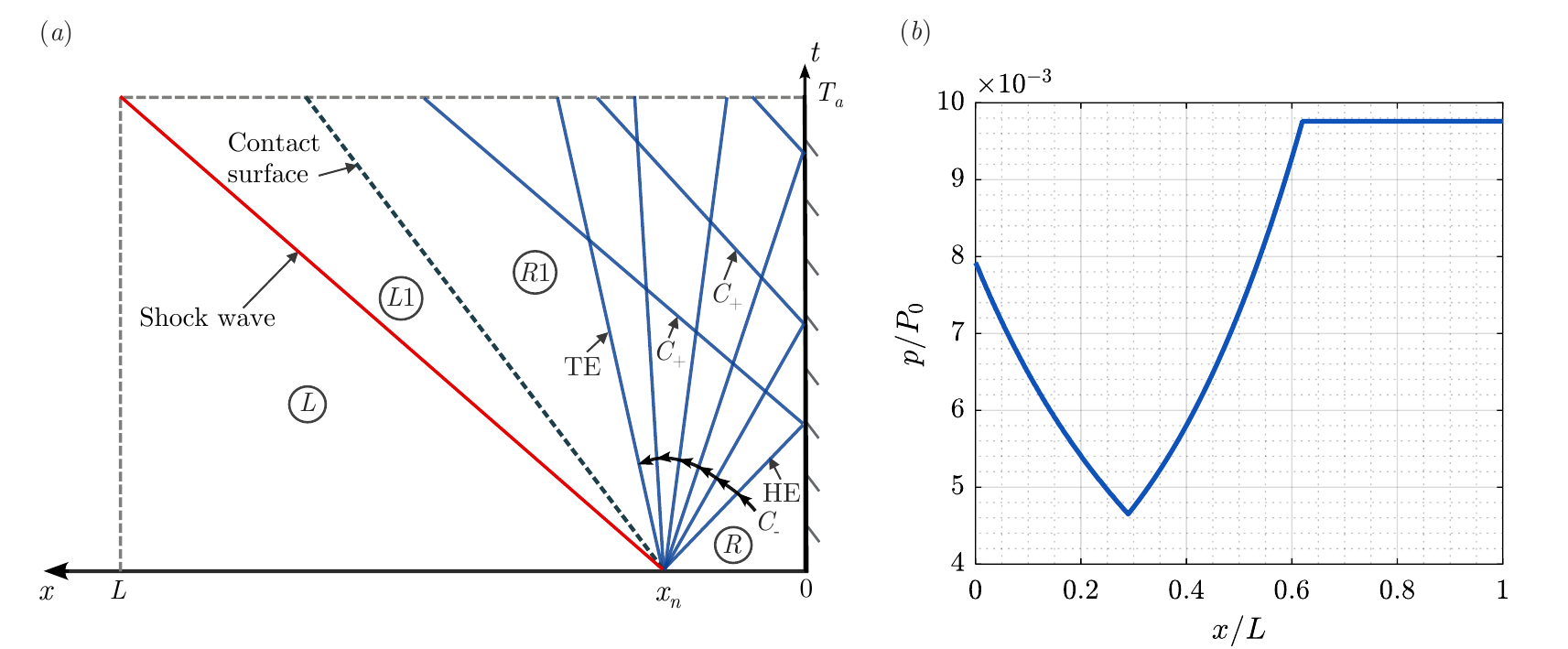}}
  \captionsetup{width=1\textwidth}
  \caption{($a$) $x$-$t$ diagram for the expansion wave reflection from the cylinder wall during the axial inflation phase. HE and TE are the head and tail of the expansion fan, respectively. $C_+$ and $C_-$ are positive and negative characteristic lines. ($b$) Pressure variation along the spike length at the end of the axial inflation phase, $t$ = $T_a$, for $\mathit{\Lambda} = 0.5$.
}
\label{fig:16}
\end{figure}

In the axial inflation phase, flow properties inside the separation region $\mathrm{S_2}$ (see figure~\ref{fig:6}$d$) change over time. Flow properties for $\mathrm{S_2}$ at the end of the axial inflation phase need to be estimated since they form the initial conditions for the radial inflation phase. The relevant terminal flow properties can be obtained by solving the conservation equations~\ref{eq:invicid_cons} with the initial conditions given by equation~\ref{eq:inf_initial} using the method of characteristics. Figure~\ref{fig:16}($a$) shows characteristic lines on the $x$-$t$ diagram, where $x = 0$ corresponds to the right-side cylinder wall and $x = L$ represents the left boundary of the separation region, \textit{i.e.}, the spike tip. A reflective boundary condition is used on the right-side wall at $x = 0$. The expansion waves propagate into region $R$ and reflect from the cylinder wall (figure~\ref{fig:16}$a$). The reflected wave then interacts with the oncoming expansion waves. Flow properties in the uniform region $R$ (ahead of the head of the expansion wave) are given by equation~\ref{eq:inf_right}. In region $R1$, behind the tail of the expansion wave, the velocity $u_{R1}$ and pressure $p_{R1}$ are given by
\begin{equation}
\begin{aligned}
u_{R1} &= u_{L1} = u_g, \\
p_{R1} &= p_{L1},
\end{aligned}
\Bigg\}
\end{equation}
where $u_{g}$ and $p_{L1}$ are given by equation \ref{eq:particle_speed} and \ref{eq:shock_strength}. The density $\rho_{R1}$ can be obtained using the isentropic relation
\begin{equation}
\rho_{R1} = \rho_R \left( \frac{p_{R1}}{p_R} \right)^{\frac{1}{\gamma}}.
\end{equation}
The solution is obtained numerically using the method of characteristics \citep{zucrow_gas_1976}. At the time instant $t$ = $T_a$ (end of the axial inflation phase), the solution for the pressure $p$ is shown in figure~\ref{fig:16}($b$) for the $\mathit{\Lambda}$ = 0.5 as a representative example. Finally, an average separation region pressure, referred to as $p_{\text{sep}}$, is defined as 
\begin{equation}
p_{\text{sep}} = \frac{1}{L} \int_{0}^{L} p \,dx.
\label{eq:sep_press}
\end{equation}
This average separation region pressure, $p_{\text{sep}}$, will provide the initial condition for the radial inflation phase.

\subsection{A model for radial inflation} \label{radial inflation model}
\begin{figure}
  \centerline{\includegraphics[width=0.6\columnwidth]{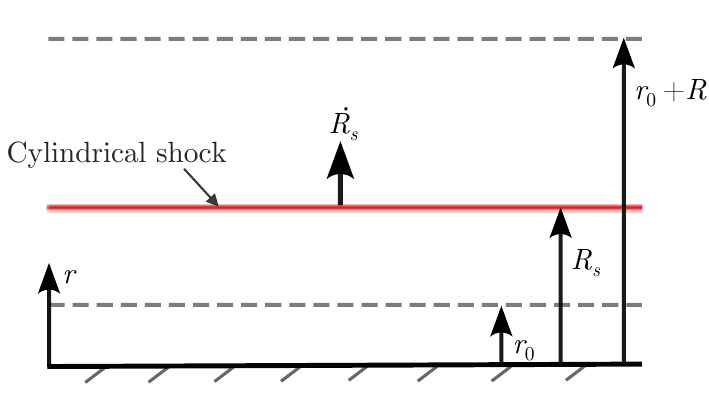}}
  \caption{Radially-propagating cylindrical shock model for the radial inflation phase. 
}
\label{fig:17}
\end{figure}

In the radial inflation phase the high-pressure gas within the separation region $\mathrm{S_2}$ expands primarily in the radial direction, causing the conical separation shock to deform and move outward from the spike axis by a radial distance approximately equal to the base-cylinder radius. To model this phase, the expansion of the high-pressure gas is assumed to be purely radial. The radial motion of the deforming shock is approximated by a simplified cylindrical shock model, in which the shock propagates radially from an initial radius $r_0$ to a final radius $r_0 + R$, where $R \left(=D/2\right)$ denotes the base cylinder radius. This process is schematically illustrated in figure~\ref{fig:17}. As the cylindrical shock propagates outward, a series of expansion waves simultaneously move inward toward the spike axis, resulting in the radial expansion of the high-pressure gas initially concentrated near the center. The duration of the radial inflation phase is estimated to be the time taken by the cylindrical shock to travel from radial location $r_0$ to $r_0 + R$. The propagation velocity of the cylindrical shock and gas properties behind the shock wave are governed by the inviscid conservation equations for mass, momentum and energy, which can be written in cylindrical coordinates $(r, \alpha,x)$ as
\begin{equation}
\frac{\partial \rho}{\partial t} + \frac{\partial (\rho u)}{\partial x} + \frac{1}{r} \frac{\partial (r \rho v)}{\partial r} + \frac{1}{r} \frac{\partial (\rho w)}{\partial \alpha}  = 0,
\end{equation}
\begin{equation}
    \frac{\partial (\rho v)}{\partial t} + \frac{\partial (\rho v u)}{\partial x} + \frac{1}{r} \frac{\partial (r \rho v^2)}{\partial r} + \frac{1}{r} \frac{\partial (\rho v w)}{\partial \alpha}  = -\frac{\partial p}{\partial r} + \frac{\rho w^2}{r},
\end{equation}
\begin{equation}
    \frac{\partial (\rho w)}{\partial t} + \frac{\partial (\rho w u)}{\partial x} + \frac{1}{r} \frac{\partial (r \rho v w)}{\partial r} + \frac{1}{r} \frac{\partial (\rho w^2)}{\partial \alpha}  = -\frac{1}{r} \frac{\partial p}{\partial \alpha} - \frac{\rho v w}{r},
\end{equation}
\begin{equation}
    \frac{\partial (\rho u)}{\partial t} + \frac{\partial (\rho u^2)}{\partial x} + \frac{1}{r} \frac{\partial (r \rho v u)}{\partial r} + \frac{1}{r} \frac{\partial (\rho w u)}{\partial \alpha}  = -\frac{\partial p}{\partial x},
\end{equation}
\begin{equation}
    \frac{\partial E}{\partial t}  + \frac{\partial}{\partial x} \left[ (E + p) u \right] + \frac{1}{r} \frac{\partial}{\partial r} \left[ r (E + p) v \right] + \frac{1}{r} \frac{\partial}{\partial \alpha} \left[ (E + p) w \right] = 0,
\end{equation}
where the total energy per unit volume is
\begin{equation}
    E = \frac{p}{\gamma - 1} + \frac{1}{2} \rho (u^2 + v^2 + w^2).
\end{equation}
Here, $u$ is the velocity in the axial direction, $v$ is the velocity in the radial direction and $w$ is the velocity in the azimuthal direction. $x$ is the axial coordinate, $r$ is the radial coordinate, $\alpha$ is the azimuthal coordinate,  $\rho$ is the density,  $p$ is the pressure and $t$ is time. The flow is assumed to be axisymmetric ($\partial/\partial \alpha = 0$) with no azimuthal velocity component ($w = 0$), and translationally invariant along the axial direction ($\partial/\partial x = 0$). Thus, the inviscid conservation equations for mass, momentum, and energy can be simplified as
\begin{equation}
\frac{\partial \rho}{\partial t} + \frac{1}{r} \frac{\partial (r \rho v)}{\partial r} = 0,
\label{eq:mass_con}
\end{equation}
\begin{equation}
    \frac{\partial (\rho v)}{\partial t} + \frac{1}{r} \frac{\partial (r \rho v^2)}{\partial r} + \frac{\partial p}{\partial r} = 0,
\label{eq:mom_r}
\end{equation}
\begin{equation}
    \frac{\partial (\rho u)}{\partial t} + \frac{1}{r} \frac{\partial (r \rho v u)}{\partial r} = 0,
\label{eq:mom_x}
\end{equation}
\begin{equation}
    \frac{\partial}{\partial t}\left[\frac{p}{\gamma - 1} + \frac{1}{2} \rho (u^2 +v^2)\right] + \frac{1}{r} \frac{\partial}{\partial r} \left[ r \left(\frac{\gamma p}{\gamma - 1} + \frac{1}{2} \rho (u^2 +v^2)\right) v \right] = 0.
\label{eq:energy}
\end{equation}
Using equation \ref{eq:mass_con} and equation \ref{eq:mom_x}, equation \ref{eq:energy} can be rewritten as 
\begin{equation}
    \frac{\partial}{\partial t}\left[\frac{p}{\gamma - 1} + \frac{1}{2} \rho v^2\right] + \frac{1}{r} \frac{\partial}{\partial r} \left[ r \left(\frac{\gamma p}{\gamma - 1} + \frac{1}{2} \rho v^2\right) v \right] = 0.
\end{equation}
With that, the governing equations for the cylindrical shockwave propagation during the radial inflation phase are given by 
\begin{equation}
\left.
\begin{aligned}
    &\frac{\partial \rho}{\partial t} + \frac{1}{r} \frac{\partial (r \rho v)}{\partial r} = 0, \\[8pt]
    &\frac{\partial (\rho v)}{\partial t} + \frac{1}{r} \frac{\partial (r \rho v^2)}{\partial r} + \frac{\partial p}{\partial r} = 0, \\[8pt]
    &\frac{\partial}{\partial t} \left[\frac{p}{\gamma - 1} + \frac{1}{2} \rho v^2 \right] 
    + \frac{1}{r} \frac{\partial}{\partial r} \left[ r \left(\frac{\gamma p}{\gamma - 1} + \frac{1}{2} \rho v^2 \right) v \right] = 0.
\end{aligned}
\quad \right\}
\label{eq:cons_wh}
\end{equation}
The conservation equation \ref{eq:cons_wh} has the exact same form as the governing equations for a cylindrical blast wave, for which an analytical solution is provided by \citet{bach1970analytical}. Following \citet{bach1970analytical}, equation \ref{eq:cons_wh} can be reduced to the following dimensionless form:
\begin{equation}
\left.
\begin{aligned}
(\phi - \xi) \frac{\partial \psi}{\partial \xi} + \psi \left( \frac{\partial \phi}{\partial \xi} \right) + \phi \left( \frac{\psi}{\xi} \right) = 2\sigma \eta \left( \frac{\partial \psi}{\partial \eta} \right), \\
(\phi - \xi) \frac{\partial \phi}{\partial \xi} + \sigma \phi + \frac{1}{\psi} \frac{\partial f}{\partial \xi} = 2\sigma \eta \left( \frac{\partial \phi}{\partial \eta} \right), \\
(\phi - \xi) \left( \frac{\partial f}{\partial \xi} - \frac{\gamma f}{\psi} \frac{\partial \psi}{\partial \xi} \right) + 2 \sigma f =
2 \sigma \eta \left( \frac{\partial f}{\partial \eta} - \frac{\gamma f}{\psi} \frac{\partial \psi}{\partial \eta} \right).
\end{aligned}
\quad \right\} 
\label{eq:cons_wh_nd}
\end{equation}
where the dimensionless parameters
\begin{align*}
    \phi(\xi, \eta) &= \frac{v(r,t)}{\dot{R}_s(t)}, \\
    f(\xi, \eta) &= \frac{p(r,t)}{\rho_{\infty} \dot{R}_s^2}, \\
    \psi(\xi, \eta) &= \frac{\rho(r,t)}{\rho_{\infty}},
\end{align*}

\begin{align*}
    \sigma(\eta) &= \frac{R_s \ddot{R}_s}{\dot{R}_s^2}, \quad \xi = \frac{r}{R_s(t)}, \quad \eta = \frac{a_{\infty}^2}{\dot{R}_s^2} = \frac{1}{M_{s}^2},
\end{align*}
are used. Here, $R_s(t)$ is the shock radius and $\dot{R}_s(t)$ is shock velocity at time $t$. $M_s$ is the shock Mach number, $\rho_{\infty}$ and $a_{\infty}$ are the density and speed of sound in the freestream, respectively. $\sigma$ is the shock decay coefficient, $\xi$ is the radial coordinate non-dimensionalized with the instantaneous shock radius $R_s(t)$ and $\eta$ is the shock strength. The boundary conditions at the shock front $\xi =1$ ($r = R_s$) are the standard Rankine-Hugoniot jump conditions, which are written as
\begin{equation}
\left.
\begin{aligned}
\phi(1,\eta) = \frac{2}{\gamma + 1}(1 - \eta),\\
f(1,\eta) = \frac{2}{\gamma + 1} - \left[\frac{\gamma - 1}{\gamma (\gamma + 1)}\right] \eta,\\
\psi(1,\eta) = \left[\frac{\gamma + 1}{\gamma - 1 + 2\eta}\right].
\end{aligned}
\quad \right\}
\end{equation}
\begin{figure}
  \centerline{\includegraphics[width=1.02\columnwidth]{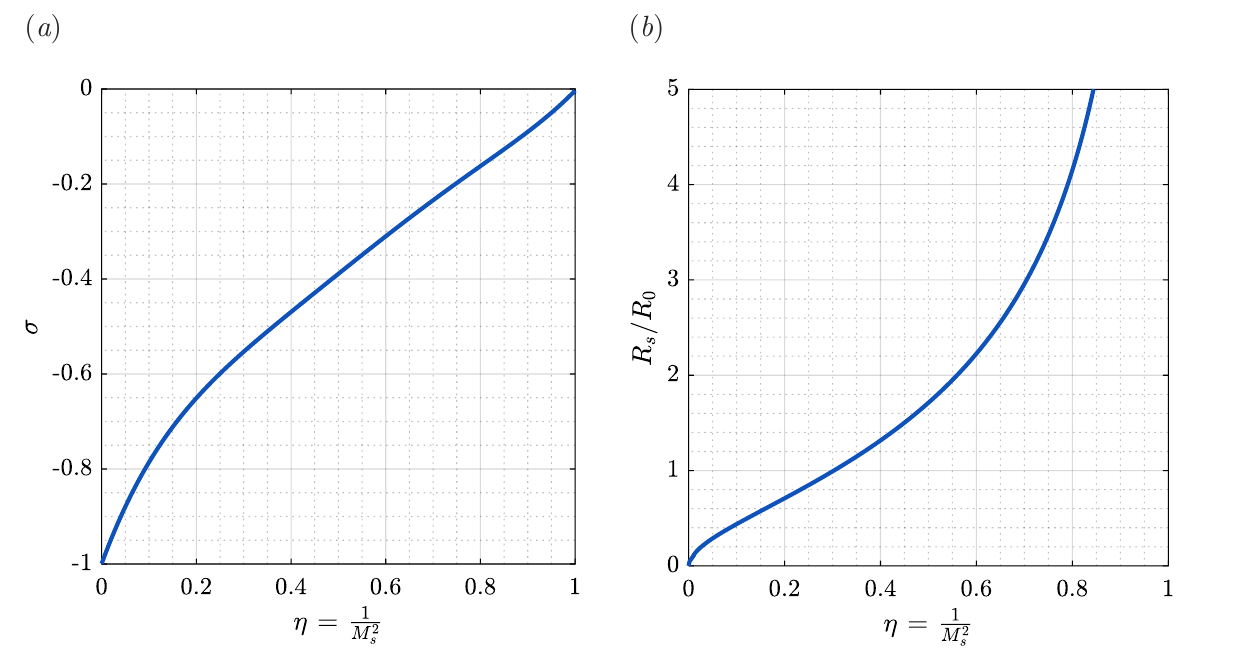}}
  \captionsetup{width=1\textwidth}
  \caption{($a$) The variation of shock decay coefficient $\sigma$ with shock strength $\eta$ for cylindrical shock waves; ($b$) the variation of shock radius $R_s/R_0$ with shock strength $\eta$ for cylindrical shock waves, $\gamma = 1.4$.
}
\label{fig:18}
\end{figure}
\citet{bach1970analytical} showed that equation \ref{eq:cons_wh_nd} can be reduced to an ordinary differential equation, the solution for which can be obtained numerically. Following their approach the solution for dimensionless parameter $\sigma(\eta)$ and shock position $R_s$ is obtained, which are shown in figures \ref{fig:18}($a$) and \ref{fig:18}($b$). In figure \ref{fig:18}($b$), $R_s$ is non-dimensionalized with the characteristic length $R_0$ given by
\begin{equation}
    R_0 = (I/2\pi \rho_{\infty} a_{\infty}^2)^\frac{1}{2},
    \label{eq:R0}
\end{equation}
where the $I$ is
\begin{equation}
    I =  \int_{0}^{r_0} \left[\frac{p}{\gamma-1} + \frac{1}{2}\rho  v^2\right]2\pi r  dr. 
\end{equation}
Here, the quantity $\frac{p}{\gamma-1} + \frac{1}{2}\rho  v^2$ is the energy per unit volume. Consequently, the integral value $I$ represents the energy per unit length, initially contained within the cylindrical region of radius $r_0$ behind the cylindrical shock. The time duration for the radial inflation phase, $T_r$, is then given by
\begin{equation}
    T_r = -\frac{R_0}{2a_{\infty}} \int_{\eta_1}^{\eta_2} \frac{(R_s/R_0)  d\eta}{\sigma \eta^{1/2}}.
\label{eq:radial inflation_time}
\end{equation}
Here, $\eta_1$ and $\eta_2$ are the initial and final shock strengths, which can be determined from the solution for $R_s/R_0$ (given in figure \ref{fig:18}$b$) corresponding to shock position $R_s = r_0$ and $R_s = r_0 + R$ respectively.
\subsubsection{Initial conditions for the radial inflation phase}
\begin{figure}
  \centerline{\includegraphics[width=0.7\columnwidth]{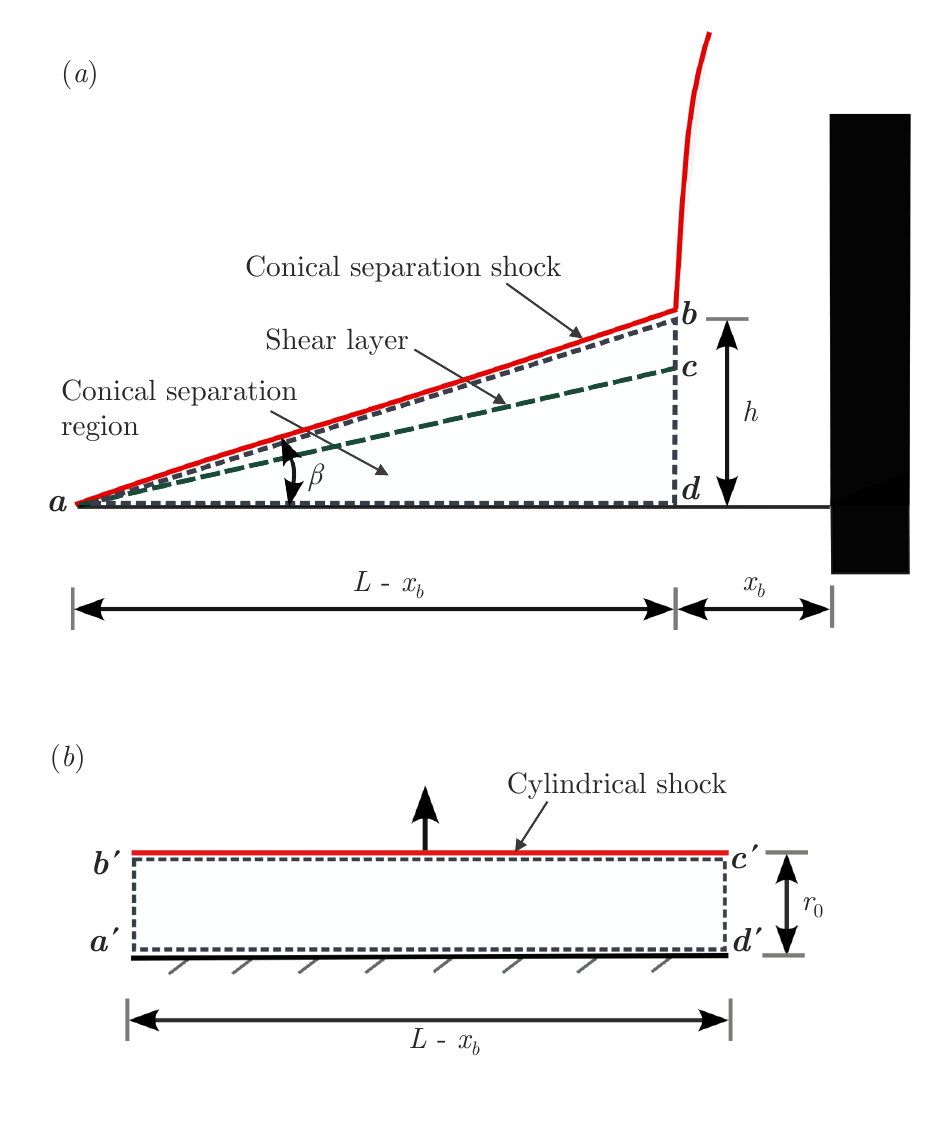}}
  \captionsetup{width=1\textwidth}
  \caption{Initial conditions for the radial inflation phase: ($a$) key flow features at the beginning of the radial inflation phase; ($b$) modeled cylindrical region $a'b'c'd'$ at the onset of the radial inflation phase. 
}
\label{fig:19}
\end{figure}

To estimate the radial inflation time $T_r$ as given by equation~\ref{eq:radial inflation_time}, the initial shock radius $r_0$ and the integral value $I$ must be prescribed. At the onset of the radial inflation phase, a conical separation shock and a conical separation region are present, as illustrated in figure~\ref{fig:19}($a$). A conical control volume $abcd$ of length $L - x_b$ and base radius $h$ is constructed to encompass both the conical separation region $acd$ and a narrow region $abc$ located downstream of the conical separation shock. This conical control volume $abcd$ is then modeled by an equivalent cylindrical control volume $a'b'c'd'$ of radius $r_0$ and the same length ($L - x_b$), as illustrated in figure~\ref{fig:19}(b). The cylindrical control volume is constructed to have the same total volume and energy as the original conical control volume. Therefore, the energy per unit length, represented by the integral value $I$, remains identical for both control volumes. The integral value $ I $ is then written as
\begin{equation}
    \begin{split}
        I &=  \int_{0}^{r_0} \left[\frac{p}{\gamma-1} + \frac{1}{2}\rho  v^2\right]2\pi r  dr \\[2ex]
        &= \frac{1}{L-x_b}\int_{a'b'c'd'} \left[\frac{p}{\gamma-1} + \frac{1}{2}\rho  v^2\right] \, dV \\[2ex]
        &= \frac{1}{L-x_b}\int_{abcd} \left[\frac{p}{\gamma-1} + \frac{1}{2}\rho  v^2\right] \, dV.
    \end{split}
\label{eq:I}
\end{equation}
To evaluate $I$, the radial velocity $v$ within the control volume $abcd$ is assumed to be negligible at the beginning of the radial inflation phase, \textit{i.e.}, $v = 0$ at $t = 0$. The pressure within the separation region $acd$ is considered to be spatially uniform and equal to the average separation region pressure $p_{\text{sep}}$, which is given by equation~\ref{eq:sep_press}. The pressure in the narrow region $abc$, situated downstream of the conical shock, is also assumed to be uniform and equal to $p_{\text{sep}}$. This assumption is justified by the expectation of pressure continuity across the shear layer which divides the separation region from region $abc$. Consequently, the pressure throughout the entire control volume $abcd$ is approximated to be uniform and equal to $p_{\text{sep}}$. Under these assumptions, the expression for the integral value $I$, given by equation~\ref{eq:I}, can be simplified as
\begin{equation}
 I = \frac{1}{L-x_b}\int_{abcd} \left[\frac{p_{\text{sep}}}{\gamma-1}\right] \, dV = \frac{1}{3}\pi h^2\left(\frac{p_{\text{sep}}}{\gamma-1}\right).
 \label{eq:I2}
\end{equation}
From figure \ref{fig:19}($a$), base radius $h$ is written as
\begin{equation}
h = (L-x_b)\tan\beta,
\end{equation}
where $\beta$ denotes the separation shock angle at the onset of the radial inflation phase. $\beta$ can be determined from the velocity of the separation shock $\mathrm{SS_2}$ during the preceding axial inflation phase.
As discussed in \S ~\ref{axial inflation model}, during the axial inflation phase, the separation shock $\mathrm{SS_2}$ propagates with a velocity $W_1$, which is given by equation \ref{eq:shock_speed}. Accordingly, the effective Mach number across the conical separation shock can be expressed as  
\begin{equation}
(M_e)_{\text{cs}} = \frac{U_{\infty} + W_1}{a_{\infty}}.
\end{equation}
Pressure downstream of the separation shock is considered as $p_{\text{sep}}$. Thus, for the given pressure ratio $p_{\text{sep}}/p_{\infty}$ and Mach number $(M_e)_{\text{cs}}$, the separation shock angle $\beta$ can be obtained by solving the Taylor–Maccoll equations for conical shock waves.\par  

Finally, an estimate of the initial shock radius $r_0$ is needed to complete this solution. The control volumes $a'b'c'd'$ and $abcd$ have equal volumes:  
\begin{equation}
\pi r_0^2(L-x_b) = \frac{1}{3}\pi h^2(L-x_b).
\end{equation}
Solving for $r_0$, we obtain  
\begin{equation}
r_0 = \frac{h}{\sqrt{3}} = \frac{(L-x_b)\tan\beta}{\sqrt{3}}.
\label{eq:r0}
\end{equation}
The initial shock radius $r_0$ obtained above is then used to estimate $\eta_1$ and $\eta_2$. Once the integral value $I$, $\eta_1$, and $\eta_2$ are determined, the time duration $T_r$ for the radial inflation phase can be estimated using equation \ref{eq:radial inflation_time}.
\begin{figure}
  \centerline{\includegraphics[width=0.9\columnwidth]{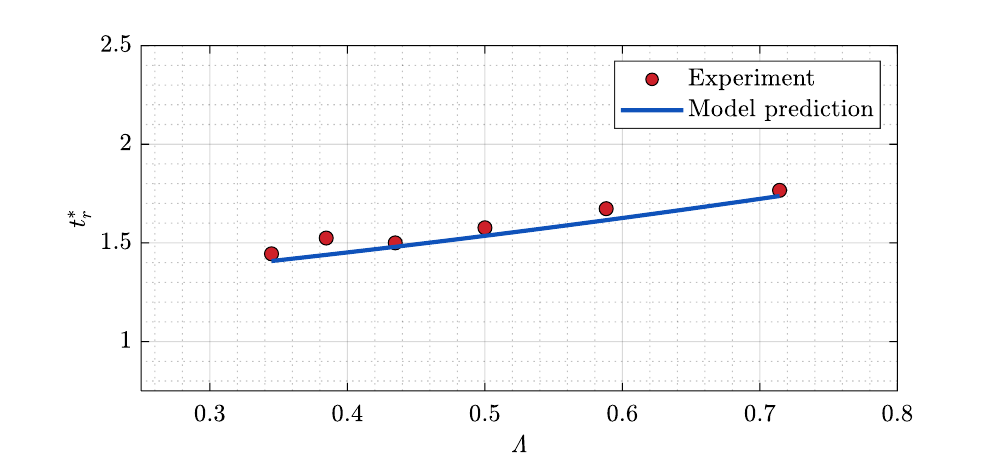}}
  \captionsetup{width=1\textwidth}
  \caption{A comparison of predicted non-dimensional radial inflation time $t^*_r \left(= \frac{T_rU_{\infty}}{D}\right)$ with experiments.
}
\label{fig:20}
\end{figure} 
\begin{figure}
  \centerline{\includegraphics[width=1.03\columnwidth]{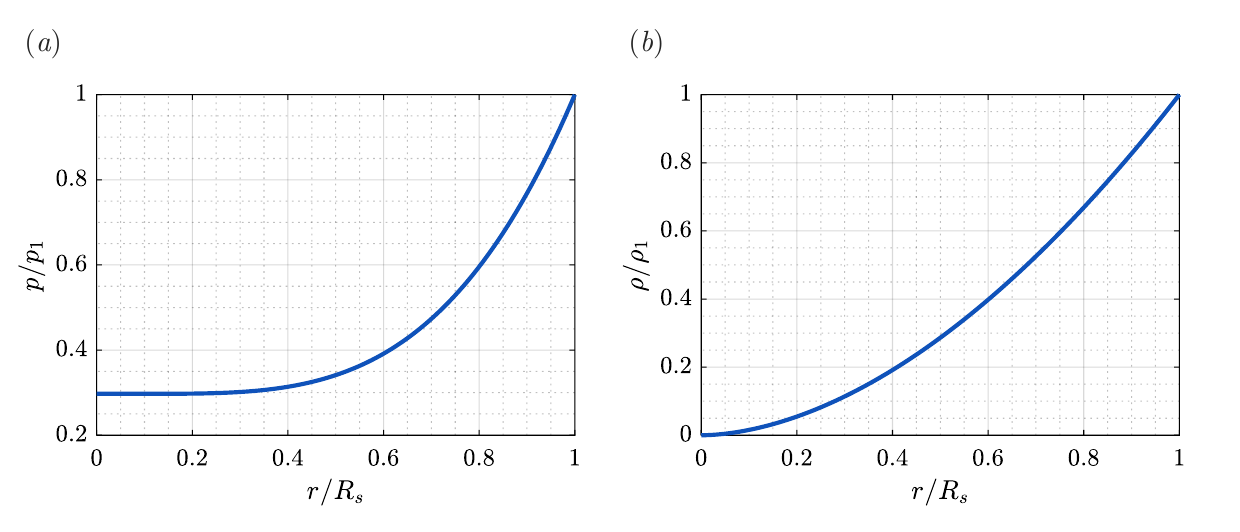}}
  \captionsetup{width=1\textwidth}
  \caption{($a$) Pressure distribution downstream of the cylindrical shock for $\mathit{\Lambda}$ = 0.5 and shock strength $\eta$ = $\eta_2$; ($b$) Density distribution downstream of the cylindrical shock for $\mathit{\Lambda}$ = 0.5 and shock strength $\eta$ = $\eta_2$. $\eta$ = $\eta_2$ corresponds to the shock radius $R_s$ = $r_0$+$R$ at the end of the radial inflation phase. $p_1$ and $\rho_1$ are the pressure and density immediately downstream of the cylindrical shock, respectively. 
}
\label{fig:21}
\end{figure}
\subsubsection{Model prediction for the radial inflation phase}
Using the above formulation, the time duration of the radial inflation phase is estimated for the six different values of $\mathit{\Lambda}$ that were studied experimentally. Figure~\ref{fig:20} presents a comparison of the non-dimensional radial inflation time, defined as $t^*_r = \frac{T_r U_{\infty}}{D}$, between experimental measurements and model predictions. The model successfully captures the trend of $t^*_r$ showing a slight increase with $\mathit{\Lambda}$, and shows good quantitative agreement with experimental measurements.

\subsubsection{Flow properties downstream of the cylindrical shock}
Flow properties downstream of the cylindrical shock are evaluated at the end of the radial inflation phase \citep{bach1970analytical}, and the results are shown in figure~\ref{fig:21} for $\mathit{\Lambda} = 0.5$ as a representative example. The pressure distribution behind the cylindrical shock wave is shown in figure~\ref{fig:21}($a$) for shock strength $\eta$ = $\eta_2$. Here we note that shock strength $\eta$ = $\eta_2$ corresponds to the shock location at the end of the radial inflation phase (shock radius is $r_0 + R$). Figure~\ref{fig:21}($b$) shows the density distribution for the same. These flow conditions provide the necessary initial conditions for modeling the collapse phase. 

\subsection{A model for collapse} \label{collapse model}
During the radial inflation phase, the radial expansion of the separation region $\mathrm{S_2}$ results in the formation of a low-pressure zone ahead of the cylinder. This low-pressure zone triggers the collapse of the shock wave system toward the cylinder. The shock wave system, comprising a normal shock ($\mathrm{NS_1}$) and a conical separation shock, resembles a bow shock which retracts toward the cylinder. At the onset of the collapse phase, a localized region of high-pressure gas forms near the spike tip. This high-pressure zone is separated from the low-pressure separation region $\mathrm{S_2}$ by an internal oblique shock, as discussed in \S\ref{collapse}. The collapse process is primarily governed by two aspects of the flow: 
\begin{enumerate}
    \renewcommand{\labelenumi}{(\roman{enumi})}
    \item the propagation of the internal oblique shock toward the cylinder through region $\mathrm{S_2}$,
    \item the subsequent motion of the bow shock as it accelerates in the same direction.\vspace{1mm}
\end{enumerate}
\par
\begin{figure}
  \centerline{\includegraphics[width=0.7\columnwidth]{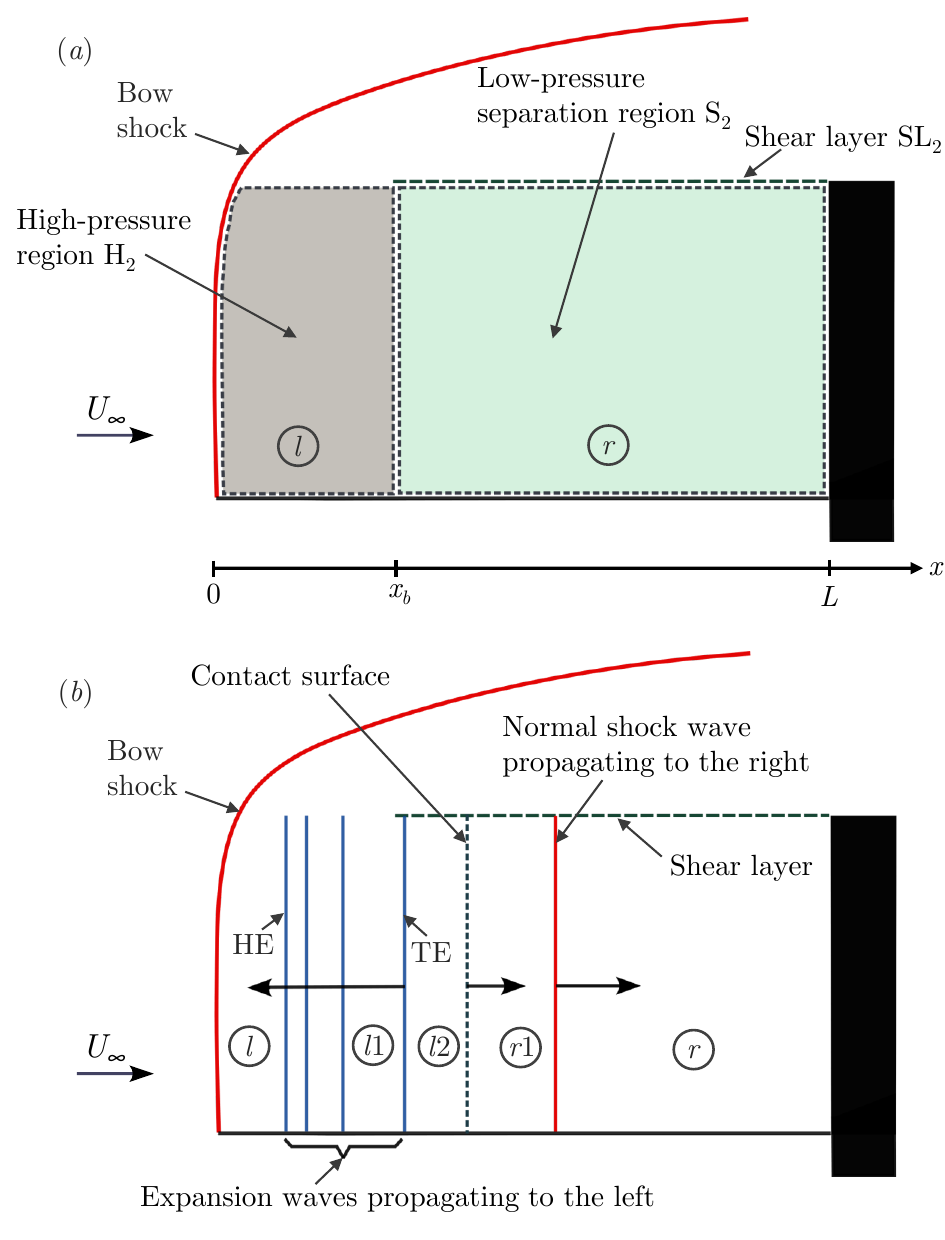}}
  \captionsetup{width=1\textwidth}
  \caption{A schematic representation of the proposed model for the collapse phase. The head and tail of the expansion fan are labeled HE and TE, respectively.  
}
\label{fig:22}
\end{figure}

A simple model of the flow is built on the basis of these observations. Figure~\ref{fig:22}($a$) shows a schematic representation of the one-dimensional flow. The high-pressure region $\mathrm{H_2}$ is situated near the spike tip, immediately downstream of the bow shock, while the low-pressure separation region $\mathrm{S_2}$ lies further downstream from the tip. Thus, the configuration consists of a high-pressure region on the left and a low-pressure region on the right. This scenario is modeled as a Riemann problem, characterized by an initial discontinuity at the interface between the two regions, similar to the formulation used for the axial inflation phase in \S\ref{axial inflation model}. The shear layer $\mathrm{SL_2}$, located above the separation region $\mathrm{S_2}$, is assumed to be parallel to the spike axis at a radial distance $R$. The flow in both $\mathrm{H_2}$ and $\mathrm{S_2}$ is assumed to be one-dimensional and aligned along the spike axis, with radial velocity components neglected for simplicity. Additionally, the flow is assumed to be initially uniform in both regions. The spatial extent in the streamwise direction of the high-pressure region $\mathrm{H_2}$ is considered as $x_b$, which corresponds to the regular bow shock stand-off distance. The governing equations remain the one-dimensional inviscid conservation laws introduced earlier in equation~\eqref{eq:invicid_cons}, with initial conditions specified as

\begin{equation}
\begin{bmatrix}
\rho \\ u \\ p 
\end{bmatrix} (x, 0) =
\begin{cases}
\begin{bmatrix}
\rho_l \\ u_{l} \\ p_l
\end{bmatrix}, & x < x_{b}, \\\\
\begin{bmatrix}
\rho_r \\ u_r \\ p_r
\end{bmatrix}, & x \geq x_{b}.
\end{cases}
\end{equation}
Here, the subscripts $l$ and $r$ denote the initial left- and right-hand states, corresponding to the high-pressure region $\mathrm{H_2}$ and the low-pressure region $\mathrm{S_2}$, respectively. The $x$-axis is aligned with the axis of symmetry, and the origin is located at the spike tip, as shown in figure~\ref{fig:22}($a$). As the flow evolves, a normal shock propagates into the low-pressure region $\mathrm{S_2}$, while a series of expansion waves travel into the high-pressure region $\mathrm{H_2}$. This is illustrated schematically in figure~\ref{fig:22}($b$). A contact surface is formed, and it moves in the same direction as the shock. The domain is divided into five different regions: $r$, the undisturbed region between the cylinder face and the shock; $r1$, between the shock and the contact surface; $l1$, between the head and tail of the expansion fan; $l2$, between the contact surface and the expansion tail; and the undisturbed region $l$ upstream of the expansion fan. While the actual collapse process involves an oblique shock, the present analysis simplifies it as a normal shock to allow for a one-dimensional modeling exercise. \par

As the expansion waves reach the spike tip and interact with the bow shock, they cause a reduction in the pressure level downstream of the shock. This weakening causes the bow shock to shift toward the cylinder. When the head of the expansion fan arrives at the spike tip, it modifies the post-shock pressure $p_1$, reducing it from its initial value $p_l$. The initial pressure $p_l$ corresponds to the post-shock pressure of a steady bow shock, which can be approximated by considering that portion of the bow shock as a steady normal shock. As the pressure behind the bow shock decreases, the effective Mach number of the incoming flow also has to decrease. This drives the bow shock to retreat toward the cylinder with a velocity $W_b$, thereby reducing the effective Mach number $M_e$, which is written as  
\begin{equation}
M_e = M_{\infty} - \frac{W_b}{a_{\infty}} \hspace{1mm},
\label{eq:bow_mach}
\end{equation}
where $a_{\infty}$ is speed of sound in freestream.
From equation \ref{eq:bow_mach}, the bow shock velocity $W_b$ is obtained to be
\begin{equation}
W_b = U_{\infty} - a_{\infty}\sqrt{\left[\frac{\gamma+1}{2\gamma} \left( \frac{p_{1}}{p_{\infty}} - 1 \right) + 1\right]}\hspace{1mm},
\label{eq:bow_speed}
\end{equation}
where $U_{\infty}$ $\left(= M_{\infty}a_{\infty}\right)$ is the freestream velocity. As the pressure downstream of the bow shock, $p_1$, continues to drop due to the interaction with expansion waves, the effective Mach number $M_e$ correspondingly decreases, while the bow shock velocity $W_b$ increases, as described by equations~\ref{eq:bow_mach} and~\ref{eq:bow_speed}. \par

\begin{figure}
  \centerline{\includegraphics[width=1\columnwidth]{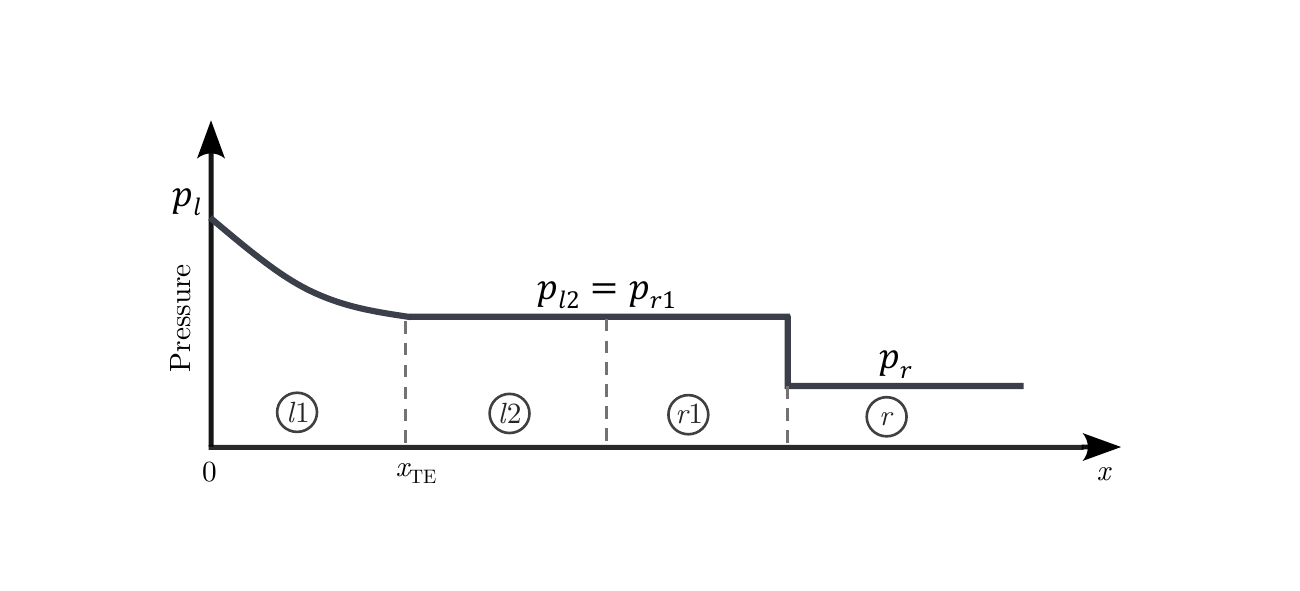}}
  \captionsetup{width=1\textwidth}
  \caption{Qualitative pressure profile along the spike during the collapse phase at the time instant where the expansion fan head meets the bow shock. 
}
\label{fig:23}
\end{figure}

Figure~\ref{fig:23} shows the qualitative pressure profile along the spike axis $x$ at the time instant when the head of the expansion fan reaches the bow shock. This pressure distribution can be obtained quantitatively by solving the inviscid conservation equations given in equation~\ref{eq:invicid_cons}. At this instant, region $l$ no longer exists, and region $l1$ begins immediately behind the bow shock. Within region $l1$ the pressure decreases monotonically along the axial direction (across the expansion waves). Following \citet{emanuel1986gasdynamics}, the pressure in region $l1$ can be obtained analytically as 
\begin{equation}
 p(x,t) = p_l\left[\frac{2}{\gamma + 1} \left(1- \frac{\gamma - 1}{2} \frac{(x-x_b)}{a_l(T_{c1} + t)}\right)\right]^\frac{2\gamma}{\gamma-1} \hspace{1mm},
 \label{eq:press_l1}
\end{equation}
where $a_l \left(= \sqrt{\gamma RT_l}\right)$ is the speed of sound in the high-pressure region $l$. $t = 0$ corresponds to the time instant when the head of the expansion wave meets the spike tip, and $x=0$ corresponds to the spike tip location. $T_{c1}$ is the time duration for the head of the expansion fan to reach the spike tip. $p(x,0)$ obtained from equation \ref{eq:press_l1} at $t = 0$ gives the pressure 
inside region $l1$. The spatial extent of the region $l1$ extends up to the tail of the expansion wave, $x_{\text{TE}}$, which is given by
\begin{equation}
x_{\text{TE}} = x_b + \left(\frac{\gamma+1}{2}u_{r1} -a_l\right)(T_{c1}+t)\hspace{1mm}. 
\label{eq:tail}
\end{equation}
Here $u_{r1}$ is the gas velocity behind the moving normal shock. For $x>x_{\text{TE}}$, a uniform region $l2$ exists, where pressure is constant at $p_{l2}$. Pressures inside regions $l2$ and $r1$ are equal since pressure is continuous across the contact surface. This gives
\begin{equation}
 p_{l2} = p_{r1}\hspace{1mm},
\end{equation}
where $p_{r1}$ is the pressure downstream of the right-moving normal shock. This pressure can be determined by solving the following implicit equation (given the initial conditions in regions $l$ and $r$)
\begin{equation}
\frac{p_l}{p_r} = \frac{p_{r1}}{p_r} \left\{ 1 - 
\frac{(\gamma - 1)(a_r / a_l)(p_{r1} / p_r - 1)}
{\sqrt{2 \gamma \left[ 2 \gamma + (\gamma + 1)(p_{r1} / p_r - 1) \right]}}
\right\}^{-\frac{2 \gamma}{\gamma - 1}}.
\end{equation}
The gas velocity $u_{r1}$ behind the moving normal shock is given by
\begin{equation}
u_{r1} = \frac{a_r}{\gamma} \left( \frac{p_{r1}}{p_r} - 1 \right)
\left( 
\frac{\frac{2\gamma}{\gamma + 1}}
{\frac{p_{r1}}{p_r} + \frac{\gamma - 1}{\gamma + 1}}
\right)^{1/2}.
\end{equation}
\par
 The bow shock continues to accelerate toward the cylinder until it reaches the tail of the expansion wave, $x_{\text{TE}}$. Beyond this, the pressure downstream of the bow shock remains constant (with no further decrease) as a uniform region exists for $x>x_{\text{TE}}$ with pressure $p_{r1}$. This causes the bow shock to then move with a constant velocity $W_2$, which can be written as    
\begin{equation}
W_2 = U_{\infty} - a_{\infty}\sqrt{\left[\frac{\gamma+1}{2\gamma} \left( \frac{p_{r1}}{p_{\infty}} - 1 \right) + 1\right]}\hspace{1mm}.
\label{eq:bow_speed4}
\end{equation}
For the rest of the collapse phase, the bow shock continues to move with the constant velocity $W_2$ until it reaches the minimum stand-off distance. \par

In the present analysis the reflected waves generated by the interaction between the bow shock and the expansion waves are neglected. The basis for this is understood by examining this interaction. A reference-frame transformation is applied such that the stationary bow shock (at the onset of the collapse phase) moves with the freestream velocity $U_\infty$, while the expansion waves overtake it from behind. By considering the portion of the bow shock as a normal shock, the interaction simplifies to a classical one-dimensional shock–rarefaction wave problem, in which a normal shock is overtaken by a rarefaction wave \citep{ben2000handbook}. Solutions to this problem indicate that the reflected waves are weak compression waves for all $\mathit{\Lambda}$ configurations considered. These weak reflections are neglected in the present model for simplicity. \par 

The time duration of the collapse phase can now be worked out as a sum of three parts:
\begin{enumerate}
    \renewcommand{\labelenumi}{(\roman{enumi})}
\item The first part is the time duration for the head of the expansion wave to reach the spike tip. The head of the expansion wave moves with the speed of sound through the quiescent gas inside the region $l$. This time duration $T_{c1}$ is given by
\begin{equation}
T_{c1} = \frac{x_b}{a_l}\hspace{1mm}.
\end{equation}
\item The second part is the time duration for the bow shock to reach the tail of the expansion fan with speed $W_b$ given by equation \ref{eq:bow_speed}. The bow shock speed $W_b$ is a function of the downstream pressure $p_1$, which in turn depends on both time $t$ and the instantaneous location of the bow shock, denoted as $x_1(t)$. The pressure distribution in the region bounded by the moving bow shock and the tail of the expansion fan ($x_1 < x < x_{\text{TE}}$) is described by equation~\eqref{eq:press_l1}. At the bow shock location $x = x_1$, the downstream pressure $p_1$ is given by
\begin{equation}
 p_1 = p_l\left[\frac{2}{\gamma + 1} \left(1- \frac{\gamma - 1}{2} \frac{(x_1-x_b)}{a_l(T_{c1} + t)}\right)\right]^\frac{2\gamma}{\gamma-1}
 \label{eq:press2_l2}.
\end{equation}
Substituting this $p_1$ into equation \ref{eq:bow_speed}, the bow shock velocity $W_b(t)$ can be rewritten as
\begin{equation}
W_b(t) = U_{\infty}
- a_{\infty}\left[
\frac{\gamma+1}{2\gamma}
\left\{
\frac{p_{l}}{p_{\infty}}
\left[
\frac{2}{\gamma + 1}
\left(
1 - \frac{\gamma - 1}{2}\frac{(x_1 - x_b)}{a_l(T_{c1} + t)}
\right)
\right]^{\frac{2\gamma}{\gamma+1}}
- 1
\right\}
+ 1
\right]^{\frac{1}{2}} .
\label{eq:bow_speed2}
\end{equation}
The time duration $T_{c2}$ required for the bow shock to reach the tail of the expansion fan can be estimated from equation~\eqref{eq:bow_speed2} using an iterative numerical approach. The location at which the bow shock meets the tail of the expansion wave is denoted by $x_2$. A detailed description of the estimation procedure is provided in Appendix~\ref{appA}.

\item The third part is the time duration $T_{c3}$ for the bow shock to reach the minimum shock standoff distance ($x_n$, measured from the cylinder face). Here, the bow shock moves with a constant speed $W_2$ given by the equation \ref{eq:bow_speed4}. At the end of the time duration, $T_{c2}$, the bow shock has already advanced a distance $x_2$ from the spike tip. The time duration $T_{c3}$ is therefore be expressed as
\begin{equation}
T_{c3} = \frac{L - x_2 - x_n}{W_2}\hspace{1mm}.
\end{equation}
The total time duration $T_c$ of the entire collapse phase is them simply written as 
\begin{equation}
T_c = T_{c1} + T_{c2} + T_{c3}\hspace{1mm}.
\end{equation}
\end{enumerate}

\subsubsection{Initial conditions for the collapse phase}

To estimate the total duration of the collapse phase, $T_c$, the initial flow conditions in regions $l$ and $r$ must be prescribed. At the onset of the collapse phase, the flow in the left side high-pressure region $l$ is assumed to be stagnant, \textit{i.e.}, $u_l = 0$. The initial pressure $p_l$ and density $\rho_l$ within this region are assumed to be uniform, and correspond to the conditions immediately downstream of the bow shock when it is steady and located at the spike tip. The segment of the bow shock upstream of region $l$ is approximated as a steady normal shock. Using the standard normal shock relations, the initial conditions in the left-side region $l$ are thus expressed as
\begin{equation}
\left.
\begin{aligned}
\rho_l &= \rho_{\infty} \left[\frac{(\gamma+1)M_{\infty}^2}{(\gamma-1)M_{\infty}^2 + 2} \right], \\
u_l &= 0, \\
p_l &= p_{\infty}\left[ 1 + \frac{2\gamma}{\gamma+1} \left( M_{\infty}^2 - 1 \right)\right].
\end{aligned}
\quad \right\}
\end{equation}
The initial conditions in the right side region $r$, representing the low-pressure separation zone $\mathrm{S_2}$, are extracted from the flow field at the end of the radial inflation phase. Specifically, these conditions correspond to the flow properties behind the cylindrical shock observed in the separation region $\mathrm{S_2}$ (as shown in figure~\ref{fig:21}). For simplicity, the properties in region $r$ are assumed uniform and are obtained by averaging the local flow quantities across the cylinder radius $R$. Additionally, consistent with the modeling assumptions, the flow is taken to be initially stagnant in this region as well, \textit{i.e.}, $u_r = 0$. The initial conditions in the right side region $r$ are thus expressed as
\begin{equation}
\begin{aligned}
\rho_r &= \frac{1}{R} \int_{0}^{R} \rho(r) \,dr, \\
u_r &= 0, \\
p_r &= \frac{1}{R} \int_{0}^{R} p(r) \,dr.
\end{aligned}
\end{equation}

\subsubsection{Model prediction for the collapse phase}
\begin{figure}
  \centerline{\includegraphics[width=0.9\columnwidth]{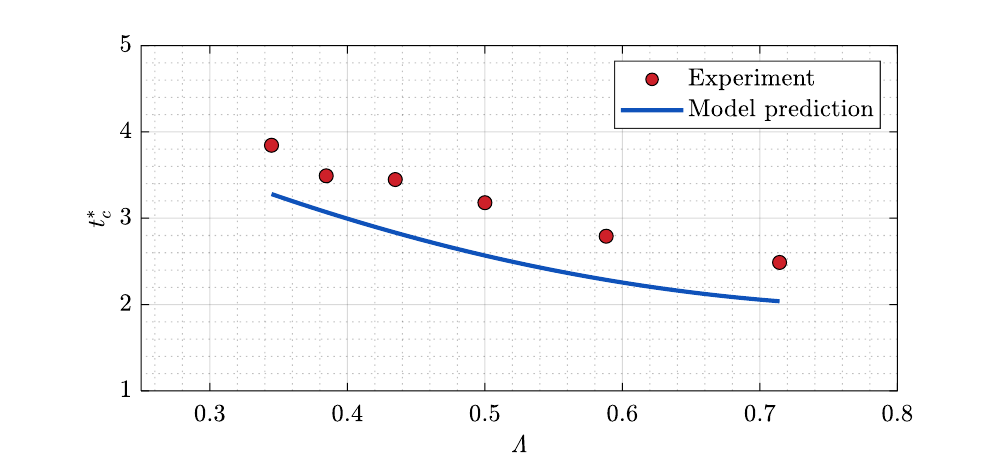}}
  \captionsetup{width=1\textwidth}
  \caption{A comparison of predicted non-dimensional collapse time $t^*_c \left(= \frac{T_cU_{\infty}}{D}\right)$ with experiments. 
}
\label{fig:24}
\end{figure} 
Using the above formulation, the time duration for the collapse phase is estimated for the six different values of $\mathit{\Lambda}$ that were studied experimentally. A comparison between the model predictions and experimental measurements of the non-dimensional collapse time, defined as $t^*_c = \frac{T_c U_{\infty}}{D}$, is presented in figure~\ref{fig:24}. While the model captures the decreasing trend of $t^*_c$ with $\mathit{\Lambda}$ seen in the experimental data, it slightly under-predicts $t^*_c$. This under-prediction arises from the choice of $x_b$ at the onset of the collapse phase, which is approximated as the regular bow-shock stand-off distance for simplicity. A more accurate estimate of the collapse time $t^*_c$ can be obtained by extracting $x_b$ directly from schlieren images (see Appendix~\ref{appB}). However, to keep the model self-contained and independent of experimental inputs, the regular bow-shock stand-off distance is used in the present formulation to provide a first-order estimate of the collapse time.
\subsection{Prediction of pulsation Strouhal number}

\begin{figure}
  \centerline{\includegraphics[width=0.9\columnwidth]{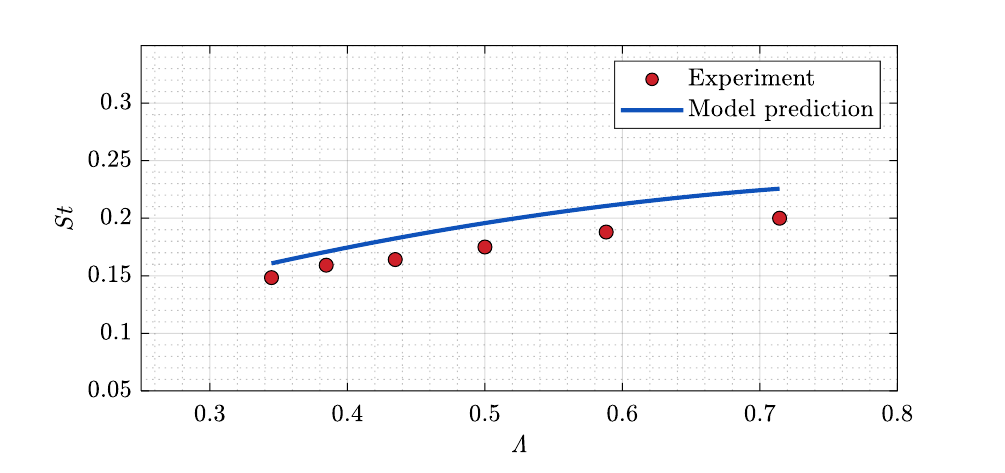}}
  \captionsetup{width=1\textwidth}
  \caption{A comparison of predicted Strouhal number for pulsation with experiment.
}
\label{fig:25}
\end{figure}

The time period $T$ for the complete pulsation cycle is simply the sum of the time durations of the three constituent phases: axial inflation, radial inflation, and collapse.
\begin{equation}
T = T_a + T_r + T_c.
\end{equation}
The corresponding pulsation frequency is $f = T^{-1}$, and the non-dimensional frequency $St$ (Strouhal number) is written as
\begin{equation}
St = \frac{fD}{U_{\infty}}.
\end{equation}
The Strouhal number estimated using the one-dimensional inviscid model is compared against experimental data in figure~\ref{fig:25}. The predicted Strouhal number from the analytical model shows good agreement with experimental results. The agreement is particularly good for lower values of $\mathit{\Lambda}$, whereas at higher $\mathit{\Lambda}$ values, a slight deviation is observed. This deviation is attributed to geometric limitations inherent in the experimental spike-cylinder configuration. As $\mathit{\Lambda}$ increases, the spike length decreases, transitioning the configuration away from an ideal double-cone geometry. At high $\mathit{\Lambda}$, the experimental model effectively resembles a triple-cone structure: a fore-cone with half-angle $\delta$, a middle-cone with zero inclination, and an aft-cone formed by the cylinder face at $90^\circ$. In contrast, the analytical model idealizes the spike body as a symmetry line representing a double-cone with $\theta_1 = 0^\circ$ and $\theta_2 = 90^\circ$, and neglects the spike tip half-angle $\delta$ and diameter $d$. This geometric disparity becomes more significant at higher $\mathit{\Lambda}$, possibly leading to the observed deviation.
Nevertheless, within the range of $\mathit{\Lambda}$ examined, the model predictions remain reasonably close to the experimental measurements. This agreement indicates that the dominant physical mechanisms driving the pulsation-type flow unsteadiness are fundamentally inviscid and are well captured by the present analytical model.
\section{Brief conclusions}
\label{conclusions}
The pulsation-type flow unsteadiness over a double-cone configuration with $\theta_1 = 0^\circ$ and $\theta_2 = 90^\circ$ was investigated using schlieren flow visualization and surface pressure measurements at Mach 6. Schlieren data clearly reveals a self-sustained, large-scale unsteadiness of the shock system, characterized by the periodic growth and collapse of the separation bubble. The non-dimensional pulsation frequency, expressed as the Strouhal number, was extracted from high-speed schlieren images using the Spectral Proper Orthogonal Decomposition (SPOD) technique. This Strouhal number closely matched the dominant frequency peak observed in the power spectral density (PSD) of surface pressure fluctuations measured at the aft-cone. The schlieren and pressure data were subject to detailed analysis to elicit the key physical mechanisms involved in the pulsation phenomenon. Though the flow dynamics associated with pulsation-type unsteadiness are inherently complex due to multiple interacting shock systems and shear layers, the leading-order flow features were identified, and a simplified picture of the pulsation cycle was constructed. The present understanding suggests that the unsteadiness is driven by the cyclic buildup of high-pressure gas near the aft-cone surface, followed by its expansion through the low-pressure separation region formed over the fore-cone. Based on these insights, a simple one-dimensional inviscid analytical model was developed to describe the pulsation cycle. The Strouhal number predicted by the model shows good quantitative agreement with experiments across the range of $\mathit{\Lambda}$ values that were studied. The model’s ability to replicate experimental trends confirms that the dominant mechanism governing the pulsation is primarily inviscid, as previously suggested by \citet{Kenworthy1978}, \citet{panaras1981pulsating}, and \citet{feszty2002utilising}. Furthermore, the present model does not corroborate the hypothesis that the supersonic jet (Edney’s jet), generated by the fore-shock and after-shock interaction (a type IV shock–shock interaction), directs high-pressure air into the separation region, which in turn drives axial and radial inflation \citep{antonov1976nonsteady, Kenworthy1978, panaras1981pulsating}. Instead, the present findings are in agreement with the interpretation of \citet{doi:10.2514/1.9034}, who argued that the gas responsible for inflating the separation bubble originates not from the annular supersonic jet but from the bubble itself. In this picture, the high-pressure gas accumulated ahead of the aft-cone during one pulsation cycle drives the expansion of the separation bubble during the subsequent cycle. This expansion of the separation bubble pushes the shock system upstream, followed by its subsequent collapse back toward the aft-cone (as discussed in \S~\ref{a model for pulsation}). In summary, the model confirms that pulsation in high-speed double-cone flows is primarily governed by inviscid dynamics. Additionally, it advances our understanding of the pulsation mechanism by clarifying the origin of the gas that drives the bubble inflation and by offering a simplified yet physically consistent description of the pulsation cycle. Overall, the model substantiates the physical interpretation of self-sustained flow pulsations derived from experimental observations.


\section*{Acknowledgements}{The authors are thankful to Dr. Vaisakh Sasidharan for the precursor experiments which aided the present work, and to M. Harish and B. M. Shiva Shankar for their assistance in the operations and maintenance of the hypersonic wind tunnel facility.}

\section*{Funding}{The authors gratefully acknowledge support for this work from a Ministry of Education (GoI) PhD Scholarship (Subhajit Das).}

\appendix
\section{}\label{appA}
During the collapse phase the bow shock accelerates as it propagates toward the cylinder until it encounters the tail of the expansion fan. The instantaneous bow shock velocity $W_b$ is expressed as 
\begin{equation}
W_b(t) = U_{\infty} - a_{\infty}\left[\frac{\gamma+1}{2\gamma} \left\{ \frac{p_{l}}{p_{\infty}}\left[\frac{2}{\gamma + 1} \left(1- \frac{\gamma - 1}{2} \frac{(x_1-x_b)}{a_l(T_{c1} + t)}\right)\right]^\frac{2\gamma}{\gamma+1} - 1 \right\} + 1\right]^\frac{1}{2},
\label{eq:bow_speed3}
\end{equation}
where the bow shock velocity depends on time $t$ and its location $x_1(t)$. To estimate the time required for the bow shock to reach the tail of the expansion fan, an iterative numerical procedure is employed.
At the initial time $t=0$, the bow shock is located at the spike tip, \textit{i.e.}, $x_1 =0$. Substituting $t=0$ and $x_1 =0$  into equation~\eqref{eq:bow_speed3} yields the initial shock velocity $W_b(0) =0$. At a subsequent time step $\Delta t$, the updated shock velocity $W_b(\Delta t)$ is computed by setting $t=\Delta t$ and $x_1=0$. Since $W_b(\Delta t)>0$, the bow shock begins to move toward the cylinder. The new shock position at time $t=\Delta t$ is given by
\begin{equation}
    x_1(\Delta t) = x_1(0) + W_b(\Delta t)\Delta t,
\end{equation}
where $x_1(0) =0$ . More generally, at time $t = n\Delta t$, the shock position is updated recursively using
\begin{equation}
    x_1(n\Delta t) = x_1\left([n-1]\Delta t\right) + W_b(n\Delta t)\Delta t,
    \label{eq:shock_loc}
\end{equation}
with $n$ denoting the current time step. Simultaneously, the tail of the expansion fan moves toward the cylinder. Its position at time $t = n\Delta t$, is given by
\begin{equation}
x_{\text{TE}}(n\Delta t) = x_b + \left(\frac{\gamma+1}{2}u_{r1} -a_l\right)(T_{c1}+n\Delta t).
\label{eq:TE}
\end{equation}
Assuming that after $m$ time steps, the bow shock meets the tail of the expansion fan at a location $x_2$. At this instant, the condition
\begin{equation}
x_1(m\Delta t) = x_{\text{TE}}(m\Delta t) = x_2
\end{equation}
holds. The time duration $T_{c2} = m\Delta t$ required for this event is thus obtained by solving equations~\eqref{eq:shock_loc} and~\eqref{eq:TE} iteratively. Once $T_{c2}$ is determined, the corresponding location $x_2$ at which the bow shock meets the tail of the expansion wave is calculated as
\begin{equation}
x_2 = x_b + \left(\frac{\gamma+1}{2}u_{r1} -a_l\right)(T_{c1}+T_{c2}). 
\end{equation}

\section{}\label{appB}
\renewcommand{\thefigure}{B\arabic{figure}}
\setcounter{figure}{0}

\begin{figure}
  \centerline{\includegraphics[width=0.9\columnwidth]{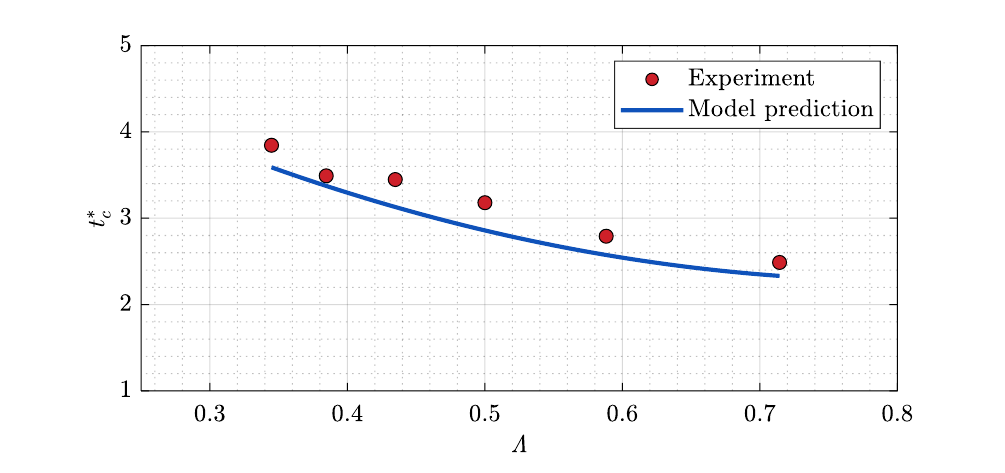}}
  \captionsetup{width=1\textwidth}
  \caption{A comparison of predicted non-dimensional collapse time $t^*_c \left(= \frac{T_cU_{\infty}}{D}\right)$ with experiments. 
}
\label{fig:B1}
\end{figure}

\begin{figure}
  \centerline{\includegraphics[width=0.9\columnwidth]{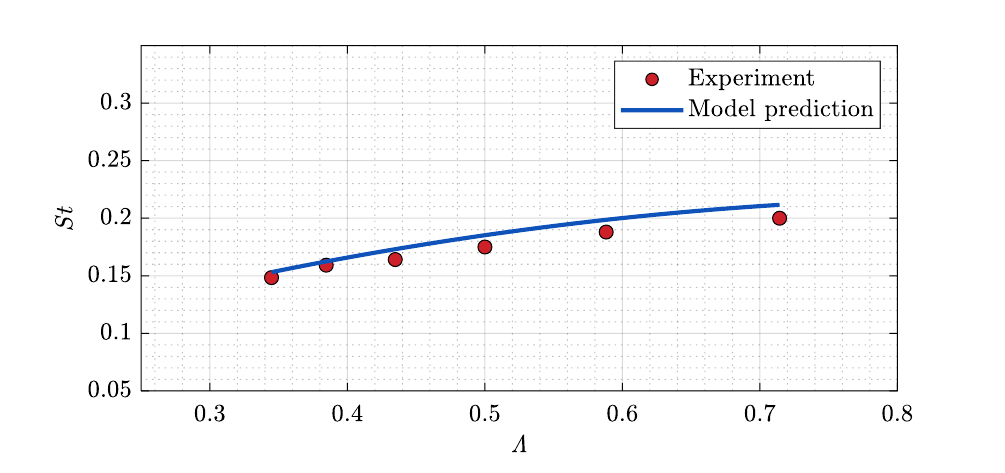}}
  \captionsetup{width=1\textwidth}
  \caption{A comparison of predicted Strouhal number for pulsation with experiment.
}
\label{fig:B2}
\end{figure}

At the onset of the collapse phase, an oblique shock separates the low-pressure separation region $\mathrm{S_2}$ from the high-pressure region that forms near the spike tip. In the present one-dimensional model, this oblique shock is represented by a normal shock generated as part of the solution to the Riemann problem formulated at this instant. The normal shock originates at a streamwise location $x_b$ from the spike tip, where the initial discontinuity of the Riemann problem is imposed. For simplicity, $x_b$ is taken to be the regular bow-shock stand-off distance (as described in \S~\ref{a model for pulsation}). A more accurate estimate of the collapse time can, however, be obtained by choosing $x_b$ as the streamwise distance from the spike tip to the mid-point of the oblique shock. This mid-point is extracted from schlieren images as the average of the streamwise positions of the leading and trailing edges of the oblique shock at the onset of collapse. This procedure was applied for all $\mathit{\Lambda}$ configurations, and the resulting non-dimensional collapse time, $t^*_c\left(= \frac{T_c U_{\infty}}{D}\right)$, is compared with experimental measurements in figure~\ref{fig:B1}. The corresponding Strouhal numbers obtained from these refined collapse times are compared with experiments in figure~\ref{fig:B2}. Although this approach yields a more accurate estimate of the pulsation Strouhal number, it relies on schlieren data. In keeping with the spirit of developing a simple and self-contained pulsation model, we refrain from using this data-assisted method and instead retain $x_b$ as the regular bow-shock stand-off distance.

\bibliographystyle{jfm}
\bibliography{jfm}

@book{Anderson2017Fundamentals,
  title={Fundamentals of Aerodynamics},
  author={Anderson, J.},
  publisher={McGraw-Hill},
  edition={6th},
  year={2016}
}

@book{anderson2003modern,
  title={Modern Compressible Flow: With Historical Perspective},
  author={Anderson, J.},
  year={2003},
  edition={3rd},
  publisher={McGraw Hill},
  isbn={0-07-242443-5}
}

@article{antonov1976nonsteady,
  title={Nonsteady supersonic flow over spiked bodies},
  author={Antonov, A. and Gretsov, V. and Shalaev, S.},
  journal={Fluid Dyn.},
  volume={11},
  pages={746--751},
  year={1976},
  publisher={Springer},
  doi={10.1007/BF01051161},
  url={https://doi.org/10.1007/BF01051161}
}

@article{bach1970analytical,
author = {Bach, G. G. and Lee, J. H. S.},
title = {An analytical solution for blast waves},
journal = {AIAA J.},
volume = {8},
number = {2},
pages = {271-275},
year = {1970},
doi = {10.2514/3.5655},
url={https://doi.org/10.2514/3.5655}
}

@book{ben2000handbook,
  title={Handbook of Shock Waves},
  author={Ben-Dor, G. and Igra, O. and Elperin, T.},
  year={2000},
  publisher={Elsevier}
}

@article{beresh2002relationship,
  title={Relationship between upstream turbulent boundary-layer velocity fluctuations and separation shock unsteadiness},
  author={Beresh, S. J. and Clemens, N. T. and Dolling, D. S.},
  journal={AIAA J.},
  volume={40},
  number={12},
  pages={2412--2422},
  year={2002},
  doi={10.2514/2.1596},
  url={https://doi.org/10.2514/2.1596}
}

@article{duvvuri2023shock,
  title={On shock-wave unsteadiness in separated flows},
  author={Duvvuri, S. and Kumar, G. and Sasidharan, V.},
  journal={S{\=a}dhan{\=a}},
  volume={48},
  number={3},
  pages={148},
  year={2023},
  publisher={Springer},
  doi={10.1007/s12046-023-02097-z},
  url={https://doi.org/10.1007/s12046-023-02097-z}
}

@techreport{edney1968anomalous,
  title={Anomalous heat transfer and pressure distributions on blunt bodies at hypersonic speeds in the presence of an impinging shock.},
  author={Edney, B.},
  number={115},
  year={1968},
  institution={Aeronautical Research Institute of Sweden}
}

@book{emanuel1986gasdynamics,
  title={Gasdynamics: Theory and Applications},
  author={Emanuel, G.},
  year={1986},
  publisher={AIAA}
}

@article{feszty2002utilising,
  title={Utilising {CFD} in the investigation of high-speed unsteady spiked body flows},
  author={Feszty, D. and Badcock, K. J. and Richards, B. E.},
  journal={Aeronaut. J.},
  volume={106},
  number={1058},
  pages={161--174},
  year={2002},
  doi={10.1017/S0001924000012173},
  url={https://doi.org/10.1017/S0001924000012173}
}

@article{doi:10.2514/1.9034,
author = {Feszty, D. and Badcock, K. J. and Richards, B. E.},
title ={{Driving mechanisms of high-speed unsteady spiked body flows. Part 1. Pulsation mode}},
journal = {AIAA J.},
volume = {42},
number = {1},
pages = {95-106},
year = {2004},
doi = {10.2514/1.9034},
URL = {https://doi.org/10.2514/1.9034},
}

@article{doi:10.2514/1.9035,
  author={Feszty, D. and Badcock, K. J. and Richards, B. E.},
  title={{Driving mechanisms of high-speed unsteady spiked body flows. Part 2. Oscillation mode}},
  journal={AIAA J.},
  volume={42},
  number={1},
  pages={107--113},
  year={2004},
  doi={10.2514/1.9035},
  url={https://doi.org/10.2514/1.9035}
}

@article{Gaitonde2015,
  author = {Gaitonde, D. V.},
  year = {2015},
  title = {Progress in shock wave/boundary layer interactions},
  journal = {Prog. Aerosp. Sci.},
  volume = {72},
  pages = {80-99},
  doi = {10.1016/j.paerosci.2014.09.002}
}

@article{Hornung_Gollan_Jacobs_2021, title={Unsteadiness boundaries in supersonic flow over double cones}, volume={916}, DOI={10.1017/jfm.2021.203}, journal={J. Fluid Mech.}, author={Hornung, H. G. and Gollan, R. J. and Jacobs, P. A.}, year={2021}, pages={A5}}

@techreport{Kenworthy1975,
  author = {Kenworthy, M. A. and Richards, B. E.},
  title = {A study of the unsteady flow over concave conic models at {Mach} 15 and 20},
  year = {1975},
  number = {AFML-TR-75-138},
  institution = {Von K\'arm\'an Institute for Fluid Dynamics},
  address = {Rhode-Saint-Gen\`ese, Belgium},
  url = {https://apps.dtic.mil/sti/tr/pdf/ADA019781.pdf}
}

@PhDThesis{Kenworthy1978,
  author  = {Kenworthy, M. A.},
  title   = {A study of unstable axisymmetric separation in high-speed flows},
  school  = {Virginia Polytechnic Institute and State University},
  year    = {1978},
  type    = {{PhD} thesis}
}

@article{Kumar_Sasidharan_Kumara_Duvvuri_2024, title={A model for frequency scaling of flow oscillations in high-speed double cones}, volume={988}, DOI={10.1017/jfm.2024.449}, journal={J. Fluid Mech.}, author={Kumar, G. and Sasidharan, V. and Kumara, A. G. and Duvvuri, S.}, year={2024}, pages={A37}}

@article{thasu2025universal, title={Aeroacoustic mechanisms explain universal behaviour in high-Mach number cylinder wakes}, volume={1010}, DOI={10.1017/jfm.2025.305}, journal={J. Fluid Mech.}, author={Thasu, P. S. and Kumar, G. and Duvvuri, S.}, year={2025}, pages={A6}}

@book{Liepmann2002Elements,
  title={Elements of Gas Dynamics},
  author={Liepmann, H. W. and Roshko, A.},
  year={2002},
  publisher={Dover Publications},
  isbn={978-0486419633}
}

@article{Mair01071952,
author = {W. A. Mair},
title = {Experiments on separation of boundary layers on probes in front of blunt-nosed bodies in a supersonic air stream},
journal = {Philos. Mag.},
volume = {43},
number = {342},
pages = {695--716},
year = {1952},
publisher = {Taylor \& Francis},
doi = {10.1080/14786440708520987},
}

@article{maull1960hypersonic,
  title={Hypersonic flow over axially symmetric spiked bodies},
  author={Maull, D. J.},
  journal={J. Fluid Mech.},
  volume={8},
  number={4},
  pages={584--592},
  year={1960},
  doi={10.1017/S002211206000074X},
  url={https://doi.org/10.1017/S002211206000074X}
}

@article{murugan2016shock,
  title={Shock wave--boundary layer interaction in supersonic flow over a forward-facing step},
  author={Murugan, J. N. and Govardhan, R. N.},
  journal={J. Fluid Mech.},
  volume={807},
  pages={258--302},
  year={2016},
  doi={10.1017/jfm.2016.617},
  url={https://doi.org/10.1017/jfm.2016.617}
}

@article{panaras1981pulsating,
  title={Pulsating flows about axisymmetric concave bodies},
  author={Panaras, A. G.},
  journal={AIAA J.},
  volume={19},
  number={6},
  pages={804--806},
  year={1981},
  doi={10.2514/3.50947},
  url={https://doi.org/10.2514/3.50947}
}

@article{panaras2009high,
  title={High-speed unsteady flows around spiked-blunt bodies},
  author={Panaras, A. G. and Drikakis, D.},
  journal={J. Fluid Mech.},
  volume={632},
  pages={69--96},
  year={2009},
  doi={10.1017/S0022112009007004},
  url={https://doi.org/10.1017/S0022112009007004}
}

@techreport{Rossiter1964Wind,
  title={Wind-tunnel experiments on the flow over rectangular cavities at subsonic and transonic speeds},
  author={Rossiter, J. E.},
  year={1964},
  institution={Aeronautical Research Council Reports \& Memoranda},
  number={ARC/R\&M-3438},
  
}

@article{sahoo2021shock,
  title={Shock-related unsteadiness of axisymmetric spiked bodies in supersonic flow},
  author={Sahoo, D. and Karthick, S. K. and Das, S. and Cohen, J.},
  journal={Exp. Fluids},
  volume={62},
  pages={89},
  year={2021},
  doi={10.1007/s00348-021-03148-x},
  url={https://doi.org/10.1007/s00348-021-03148-x}
}

@article{sasidharan2021large,
  title={Large-and small-amplitude shock-wave oscillations over axisymmetric bodies in high-speed flow},
  author={Sasidharan, V. and Duvvuri, S.},
  journal={J. Fluid Mech.},
  volume={913},
  pages={R7},
  year={2021},
  publisher={Cambridge University Press},
  doi={10.1017/jfm.2021.115},
  url={https://doi.org/10.1017/jfm.2021.115}
}

@article{thasu2022strouhal,
  title={Strouhal number universality in high-speed cylinder wake flows},
  author={Thasu, P. S. and Duvvuri, S.},
  journal={Phys. Rev. Fluids},
  volume={7},
  pages={L081401},
  year={2022},
  doi={10.1103/PhysRevFluids.7.L081401},
  url={https://doi.org/10.1103/PhysRevFluids.7.L081401}
}

@article{Thasu2024,
  title={Measurement of freestream noise in a hypersonic wind tunnel},
  author={Thasu, P. S. and Duvvuri, S.},
  journal={Exp. Fluids},
  volume={65},
  pages={45},
  year={2024},
  doi={10.1007/s00348-024-03783-3},
  url={https://doi.org/10.1007/s00348-024-03783-3}
}

@article{touber2011low,
  title={Low-order stochastic modelling of low-frequency motions in reflected shock-wave/boundary-layer interactions},
  author={Touber, E. and Sandham, N. D.},
  journal={J. Fluid Mech.},
  volume={671},
  pages={417--465},
  year={2011},
}

@article{towne2018spectral,
  title={Spectral proper orthogonal decomposition and its relationship to dynamic mode decomposition and resolvent analysis},
  author={Towne, A. and Schmidt, O. T. and Colonius, T.},
  journal={J. Fluid Mech.},
  volume={847},
  pages={821--867},
  year={2018}, 
}

@article{wood1962hypersonic,
  title={Hypersonic flow over spiked cones},
  author={Wood, C. J.},
  journal={J. Fluid Mech.},
  volume={12},
  number={4},
  pages={614--624},
  year={1962},  
}

@article{doi:10.1260/1475472054771367,
  author={Zapryagaev, V. I. and Kavun, I. N.},
  title={Experimental Investigation of Self-Sustained Oscillations on the Spike-Tipped Cylinder in Supersonic Flow},
  journal={Int. J. Aeroacoust.},
  volume={4},
  number={3},
  pages={363--372},
  year={2005},
  doi={10.1260/1475472054771367},
  url={https://doi.org/10.1260/1475472054771367}
}

@book{zucrow_gas_1976,
  author={Zucrow, M. J. and Hoffman, J. D.},
  title={Gas Dynamics, Vol.~1},
  publisher={John Wiley \& Sons},
  year={1976},
  isbn={9780471996902}
}

\end{document}